# Materials challenges for SrRuO$_3$: from conventional to quantum electronics

M. Cuoco[1]*, A. Di Bernardo[2]†

[1]CNR-SPIN, c/o University of Salerno, I-84084 Fisciano, Salerno, Italy
[2]University of Konstanz, Department of Physics, 78457 Konstanz, Germany

Correspondence to:
Mario Cuoco (mario.cuoco@spin.cnr.it)
Angelo Di Bernardo (angelo.dibernardo@uni-konstanz.de)

**Abstract**

The need for faster and more miniaturised electronics is challenging scientists to develop novel forms of electronics based on quantum degrees of freedom different from electron charge. In this fast-developing field, often referred to as quantum electronics, the metal-oxide perovskite SrRuO$_3$ can play an important role thanks to its diverse physical properties, which have been intensively investigated, mostly for conventional electronics.

In addition to being chemically stable, easy to fabricate with high quality and to grow epitaxially onto many oxides – these are all desirable properties also for conventional electronics – SrRuO$_3$ has interesting properties for quantum electronics like itinerant ferromagnetism and metallic behaviour, strong correlation between magnetic anisotropy and spin-orbit coupling, strain-tuneable magnetization, anomalous Hall and Berry effects.

In this Research Update, after describing the main phenomena emerging from the interplay between spin, orbital, lattice and topological quantum degrees of freedom in SrRuO$_3$, we discuss the challenges still open to achieve control over these phenomena. We then provide our perspectives on the most promising applications of SrRuO$_3$ for devices for conventional and quantum electronics. We suggest new device configurations and discuss the materials challenges for their realization. For conventional electronics, we single out applications where SrRuO$_3$ devices can bring competitive advantages over existing ones. For quantum electronics, we propose devices that can help gain a deeper understanding of quantum effects in SrRuO$_3$ to exploit them for quantum technologies. We finally give an outlook about properties of SrRuO$_3$ still waiting for discovery and applications that may stem from them.

# 1. Introduction

The interest of the research community in SrRuO$_3$ has been kept high for almost 60 years[1,2], as result of the coexistence of its fascinating physical properties with the easiness of its fabrication and integration in oxide heterostructures and devices.

Despite the intense research activity done on SrRuO$_3$ and SrRuO$_3$-based heterostructures, new physical properties and applications of SrRuO$_3$ are continuously being discovered. SrRuO$_3$ combines a range of interesting properties including good metallic conductivity at low temperatures ($T$s), magnetic ordering with perpendicular magnetic anisotropy, narrow domain walls, strong spin-orbit coupling strength[2]. In addition to this rich physics, another advantage of SrRuO$_3$ for device applications is that most of the SrRuO$_3$ properties can be modulated. The possibilities to tune these properties are many and include changes in the SrRuO$_3$ thickness and stoichiometry, strain application and interfacing of SrRuO$_3$ to other oxides in heterostructures and superlattices[2].

Several review articles have been written on SrRuO$_3$ over the years including a very comprehensive review[2] on SrRuO$_3$ properties and applications. In addition to several papers[3-6] summarizing the main results reported in the literature on the anomalous Hall effect (AHE) and topological Hall effect (THE) in SrRuO$_3$, we are aware of another review article[7] recently published which describes the main applications of SrRuO$_3$-based heterostructures.

The aim of this work is to put the results obtained to date on SrRuO$_3$ in perspectives and discuss which materials challenges have to be addressed to realise SrRuO$_3$-based devices with better performance and novel functionalities compared to existing ones. In addition to analysing these challenges, we propose specific examples of electronic devices with corresponding geometries that have never been realised to date. The fabrication and testing of these devices can serve as a stimulus to the research community not only from an application-related perspective, but also to gain a better understanding of quantum phenomena recently discovered in SrRuO$_3$. We propose, for example, devices that would allow to differentiate between real-space or momentum-space contributions to the SrRuO$_3$ Berry curvature. Differentiating between these contributions is a key step to engineer future quantum devices exploiting AHEs and THEs in SrRuO$_3$ for their functioning.

In section 1 of this review, we describe the main physical properties of SrRuO$_3$, and we report the deposition techniques and methodologies that can be used to fabricate SrRuO$_3$ devices for technological applications. We highlight in particular techniques that are not only suitable to produce devices with optimal properties but also with high reproducibility and scalability. After discussing the structural parameters and mechanisms that mostly affect the physical properties of SrRuO$_3$, we review progress made on the fabrication of free-standing SrRuO$_3$ structures. We also review how SrRuO$_3$ properties change when the SrRuO$_3$ dimensionality is lowered from the three-dimensional (3D) to the zero-dimensional (0D) regime and quantum effects become increasingly more relevant.

In section 2, we consider the most promising applications of SrRuO$_3$ for both conventional and quantum electronics and propose possible devices that can be made for each type of applications. Whilst describing these electronic devices and discussing possible layouts for their realization, we also outline the materials challenges that have to be addressed for their realization.

For conventional electronics, we focus on applications for which SrRuO$_3$-based devices would offer a competitive advantage over existing devices. The first class of applications include room-$T$ spintronic devices and cryogenic memories, where two distinct properties of SrRuO$_3$ namely its high spin-orbit coupling and narrow domain walls are used, respectively, to make devices that can offer better performance than existing ones. For the second class of applications, we suggest exploiting the properties of freestanding SrRuO$_3$ membranes under strain to realise nanoelectromechanical systems with unprecedentedly high figures of merit.

For quantum electronics, we focus on effects related to the non-trivial Berry curvature of SrRuO$_3$ and suggest the realization of novel devices, where real-space and momentum-space contributions to Hall signals can be differentiated and separately manipulated. We also propose new schemes of superconducting devices, where SrRuO$_3$ is coupled to a superconductor. The SrRuO$_3$-based superconducting devices that we suggest can host topological superconductivity or spin-polarised superconducting currents – which can be reversibly modulated by tuning the SrRuO$_3$ Berry curvature.

## 1.1 Main properties and growth techniques

In this section, we review the main physical properties of SrRuO$_3$ including its structural, electronic transport and magnetic properties. We list some parameters and typical values that can be used as benchmark comparison to evaluate the degree of quality of SrRuO$_3$ samples. We then discuss which growth techniques appear most promising to date for the reliable fabrication of SrRuO$_3$ thin films with optimal parameter values (i.e., close to bulk) and over large scale. The growth of SrRuO$_3$ thin films with properties identical to bulk is essential to investigate emergent phenomena and discover new quantum effects in SrRuO$_3$. A high scalability in making optimal SrRuO$_3$ thin films is in turn crucial for the development of device applications exploiting such effects and phenomena.

SrRuO$_3$ is a layered oxide perovskite of the ABO$_3$ type belonging to the Ruddlesden-Popper series of ruthanates, Sr$_{n+1}$Ru$_n$O$_{3n+1}$, with number of layers $n = \infty$. As for several other ABO$_3$ perovskites, the unit cell of bulk SrRuO$_3$ has an orthorhombic crystal symmetry at room $T$ (space group *Pbnm*). In bulk single-crystal form, SrRuO$_3$ undergoes a structural transition first into a tetragonal phase (space group *I4/mcm*) as $T$ is increased to 547 °C and then into a cubic phase (space group *Pm3m*) as $T$ is further increased up to 677 °C (ref. [2]).

In the unstrained orthorhombic phase at room $T$, the Ru-O bond is about 2 times shorter in length than the Sr-O bond which introduces a distortion of the RuO$_6$ octahedra. The distortion of the RuO$_6$ octahedra, which can be manipulated via strain engineering, is a key structural parameter affecting some of the SrRuO$_3$ physical properties, as further discussed below in this review. The lattice parameters of the orthorhombic cell (space group *Pbnm*) are[2,8] $a_{or} = 5.57$ Å, $b_{or} = 5.53$ Å, and $c_{or} = 7.85$ Å. The orthorhombic unit cell consists of four units of the ideal cubic perovskite structure, which results in a pseudocubic lattice constant[2] $a_{pc} = 3.93$ Å (Fig. 1). We note that throughout the review we use the subscripts 'or' and 'pc' to refer to the orthorhombic and pseudocubic unit cell parameters, respectively.

SrRuO$_3$ was reported as the first oxide exhibiting ferromagnetism[1] due to itinerant electrons below a Curie temperature ($T_{Curie}$) of ~ 160 K (ref. [2]), it has a relatively high saturation moment of 1.6 $\mu_B$/Ru atom[9,10] at $T = 0$ ($\mu_B = 9.27 \times 10^{-24}$ J·T$^{-1}$ being the Bohr magneton) and it usually exhibits perpendicular magnetic anisotropy when epitaxially grown as thin film under compressive strain onto a (001) SrTiO$_3$ substrate[2]. This magnetic anisotropy, however, can change depending on substrate-induced strain and orientation (see also section 1.2). The $T$-dependence of the SrRuO$_3$ electronic transport properties also shows that SrRuO$_3$ has very good metallicity at low $T$s due to its Fermi liquid behavior[11] for $T < 10$ K (metallicity is defined from the slope d$\rho$/d$T$ of the resistivity $\rho$ versus $T$ curve). As $T$ is increased and approaches room $T$, metallicity in SrRuO$_3$ progressively gets worse[12].

In addition to being one of the metal oxides with the lowest electrical resistivity at room $T$ ($\rho < 200$ $\mu\Omega$·cm at $T = 300$ K; ref. [13]), SrRuO$_3$ is also thermally stable and chemically inert[2, 14 - 16] and it exhibits very good lattice matching to other functional oxides[2,17] including piezoelectrics, e.g., Pb(Zr,Ti)O$_3$ or (Ba,Sr)TiO$_3$ (refs. [18,19]). These are some of the reasons why SrRuO$_3$ has been widely exploited to date both as metallic substrate for the growth of complex oxide heterostructures[2,18-22] and as ferromagnet in spintronic and superconducting devices including magnetic tunnel junctions[23-25] and Josephson junctions[26-28].

Several growth techniques have been used over the years to grow SrRuO$_3$ with the above-listed properties. The vast majority of the studies to characterise these properties of SrRuO$_3$ has been carried out on SrRuO$_3$ thin films. Bulk single crystals of SrRuO$_3$ are difficult to grow, and this is the main reason why the physical properties of SrRuO$_3$ have been mostly investigated in its thin film form[2]. The growth of bulk single crystals of SrRuO$_3$ by the floating zone technique, which is the preferred method to synthesize single crystals with low levels of disorder, is made difficult by the large amount of RuO$_2$ that evaporates during the SrRuO$_3$ growth. In general, obtaining good-quality single crystals is challenging for any ruthenates because the pseudo-binary diagram (SrO-RuO$_2$) is open. This open phase diagram implies that, at a given $T$ and for fixed concentrations of the SrO and RuO$_2$, it is not possible to determine the composition of the final compound resulting from growth. The composition in fact changes due to the Ru evaporation that constantly occurs during growth.

For a bulk single-crystal of the Ruddlesden-Popper series Sr$_{n+1}$Ru$_n$O$_{3n+1}$, the RuO$_2$ mass loss during the floating-zone growth is proportional to the $n$ value of the compound within the series[29] ($n = \infty$ for SrRuO$_3$). The evaporated RuO$_2$ accumulates over the walls of the quartz tube of the image furnace during growth. The accumulated RuO$_2$ reduces the intensity of the infrared light in the molten zone, which makes the growth extremely unstable[29]. This is the main reason why, although the floating zone technique has been routinely used to grow high-quality crystals of other compounds in the $Sr_{n+1}Ru_nO_{3n+1}$ Ruddlesden-Popper series[30,31] like the unconventional superconductor $Sr_2RuO_4$ ($n = 1$) and the metamagnet $Sr_3Ru_2O_7$ ($n = 2$), it has not been extensively used for $SrRuO_3$ ($n = \infty$) single crystals. Other factors are also crucial to get good-quality $SrRuO_3$ single crystals, which contribute to make the growth process very challenging. These crucial factors include the high quality of the feed rod and the excess $RuO_2$ amount, which has to be added to the rod before growth to compensate for Ru losses.

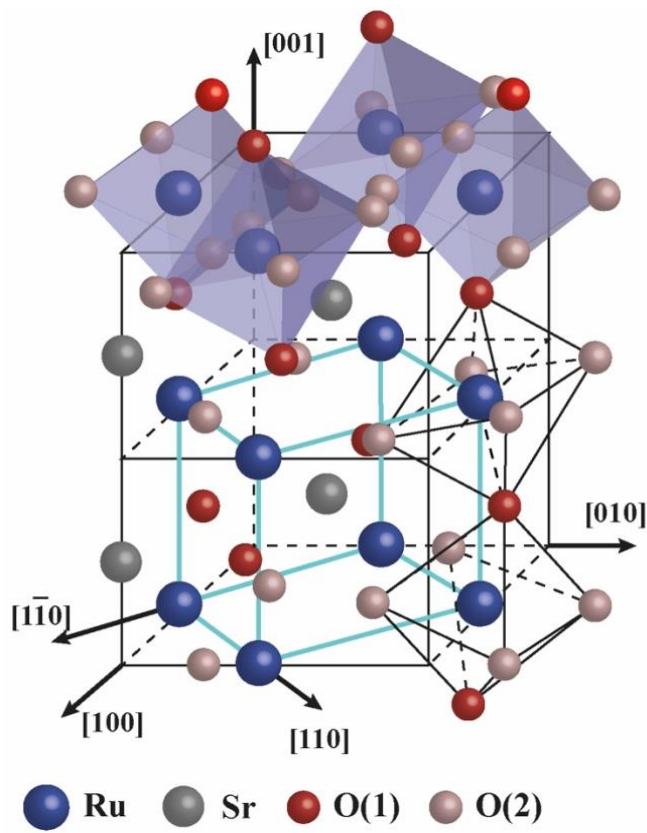

**FIG. 1. Crystallographic structure of SrRuO$_3$.**
**Illustration of the orthorhombic unit cell of SrRuO$_3$ characterized by tilt of the RuO$_6$ octahedra (in purple). The Sr, Ru, apical O and planar O atoms are represented with grey, blue, red, pink spheres, respectively.**
**The pseudocubic unit cell is also shown (light blue box) together with representative crystallographic axes of the orthorhombic unit cell which are indicated by arrows. [Figure drawn based on ref. [8]].**

$SrRuO_3$ single crystals of good quality grown by the floating zone technique have been obtained thanks to the installation of a cold trap[29]. The cold trap allows the evaporated $RuO_2$ to collect onto the trap surface other than on the walls of the quartz tube.

Unlike for single crystals, the growth of epitaxial $SrRuO_3$ thin films onto lattice-matched substrates is relatively easy to carry out and $SrRuO_3$ thin films of very high quality have been obtained by many groups using a variety of chemical and physical deposition techniques[32]. These deposition techniques include 90° off-axis magnetron sputtering[33,34], reactive evaporation[12,35,36], metalorganic chemical vapor deposition[37], chemical solution deposition[38], pulsed laser deposition (PLD) [14,39-44], and molecular beam epitaxy[13,32,45-48] (MBE). A comparison of the residual resistivity ratio (RRR), which provides quantitative information on the number of crystallographic defects and impurities inside a material, for $SrRuO_3$ thin films grown with the above techniques has been reported in ref.[32] and it is shown in Fig. 2a. The highest RRR values obtained to date are above 80 for $SrRuO_3$ thin films (grown by MBE[13,48] or reactive evaporation[35] on (001) $SrTiO_3$ substrates) and of ~ 192 for bulk $SrRuO_3$ single crystals[29].

$SrRuO_3$ thin films epitaxially grown with the above techniques usually have different lattice parameters compared to bulk because of epitaxial strain induced by the substrate. As recently observed in ref.[49], $SrRuO_3$ thin films with RRR higher than 50 have mostly been obtained on growth substrates having a small lattice mismatch with $SrRuO_3$ such as (001) $SrTiO_3$ (refs.[13,35,48-52]) and (110) $DySrO_3$ (refs. [53,54]). The lattice mismatch for $SrRuO_3$ is of ~ -0.6% with (001) $SrTiO_3$ and of ~ 0.4% with (110) $DySrO_3$. We note here that epitaxial $SrRuO_3$ thin films grown onto a (001) $SrTiO_3$ substrate usually have a tetragonal structure (*4mmm* space group) for small thicknesses (up to 4-6 nm), and a monoclinic structure (*P2₁/m* space group) for larger thicknesses[55,56]. Following the conventional notation, in the literature this monoclinic structure with the angle γ (close to 90°) between the [100]$_{or}$ and [001]$_{or}$ axes is also denoted as orthorhombic[55].

The structural transition from tetragonal to orthorhombic is correlated to a change in the $RuO_6$ octahedra tilting (see section 1.2), although the origin of this change with thickness remains unclear[49]. Using low-energy electron diffraction and high-resolution scanning transmission electron microscopy, it has been shown[57] that, unlike for other oxide perovskites, the $RuO_6$ octahedra tilting is already present in one-unit-cell-thick $SrRuO_3$ on (001) $SrTiO_3$.

In addition to a very high RRR[51,52,58] (> 50), there are several other physical properties that can be regarded as hallmark signatures of high quality for $SrRuO_3$ thin films. Indications of high $SrRuO_3$ thin film quality include low residual $\rho$ at liquid helium $T$ (~ 4.2 K), a high $T_{Curie}$, a strong perpendicular magnetic anisotropy (for thin films grown under compressive strain), and a low in-plane mosaic spread.

Low residual $\rho$ is an indication of low concentration of defects and of a good stoichiometry. Ultra-high quality $SrRuO_3$ thin films have residual resistivity lower than 3 $\mu\Omega$ cm at $T$ = 4.2 K, which is consistent with their very high RRR values[13,48,49,51,52]. A high $T_{Curie}$ is also a signature of good stoichiometry, since Ru deficiencies are one of the main reasons for a decrease in $T_{Curie}$ (see section 1.2). The highest $T_{Curie}$ reported for $SrRuO_3$ thin films deposited on (001) $SrTiO_3$ is of ~ 152 K (ref.[53]). For tensile-strained $SrRuO_3$ thin films grown on (110) $DySrO_3$ substrates, $T_{Curie}$ as high as 169 K have been instead measured[35,54].

Apart from being desirable for spintronics applications, strong perpendicular magnetic anisotropy is also a signature of high crystallinity in compressive-strained epitaxial $SrRuO_3$ thin films, since it is normally lowered by grain boundaries and other defects[59,60]. The $SrRuO_3$ thin film in ref.[48] do not only have the highest RRR (~ 86) reported to date, but they are also the first to show single-domain perpendicular magnetization. The single-domain perpendicular magnetization of the thin films in ref.[48] is evidenced by the fact that their remanent magnetization to saturation magnetization ratio (i.e., squareness) is of ~ 0.97. Single-domain perpendicular magnetization in $SrRuO_3$ is a desirable property for spintronics[61], and its recent realization in ref.[48] will certainly contribute to further applications of $SrRuO_3$ in oxide spintronics at cryogenic $T$s.

Low in-plane mosaic spread is also a good indication of high thin film quality. The in-plane mosaic spread can be estimated for $SrRuO_3$ by measuring the full width at half maximum, FWHM, for the rocking curves of the $(001)_{pc}$ or $(002)_{pc}$ peaks. In Fig. 2b, we reproduce a figure from ref.[32], where the authors compare the FWHM of the $(001)_{pc}$ and $(002)_{pc}$ peaks of $SrRuO_3$ thin films grown with different deposition techniques and substrates. The data in Fig. 2b show that SrRuO$_3$ thin films with very low amount of in-plane mosaic spread (i.e., FWHM ≤ 0.01°) have been obtained by several groups using either PLD or MBE growth.

Apart from targeting the above-listed parameter values, which are good indicators of high SrRuO$_3$ thin film quality, another main challenge to address for future applications of SrRuO$_3$ in conventional and quantum electronics is to understand how to scale up SrRuO$_3$ thin films and in turn devices based on them. The scaling up therefore implies not only growing epitaxial SrRuO$_3$ thin films with optimal physical properties, but also doing this over large areas and pattern then the films into devices with reliable functioning.

Optimising the growth of high-quality epitaxial SrRuO$_3$ thin films on Si, the material at the core of complementary metal-oxide semiconductor (CMOS) technology, over areas comparable to the size of Si wafers used by the semiconductor industry (> 4'' in diameter) can lead, for example, to the integration of the fabrication of SrRuO$_3$-based devices into the industrial processes and fabs of the semiconductor industry[32].

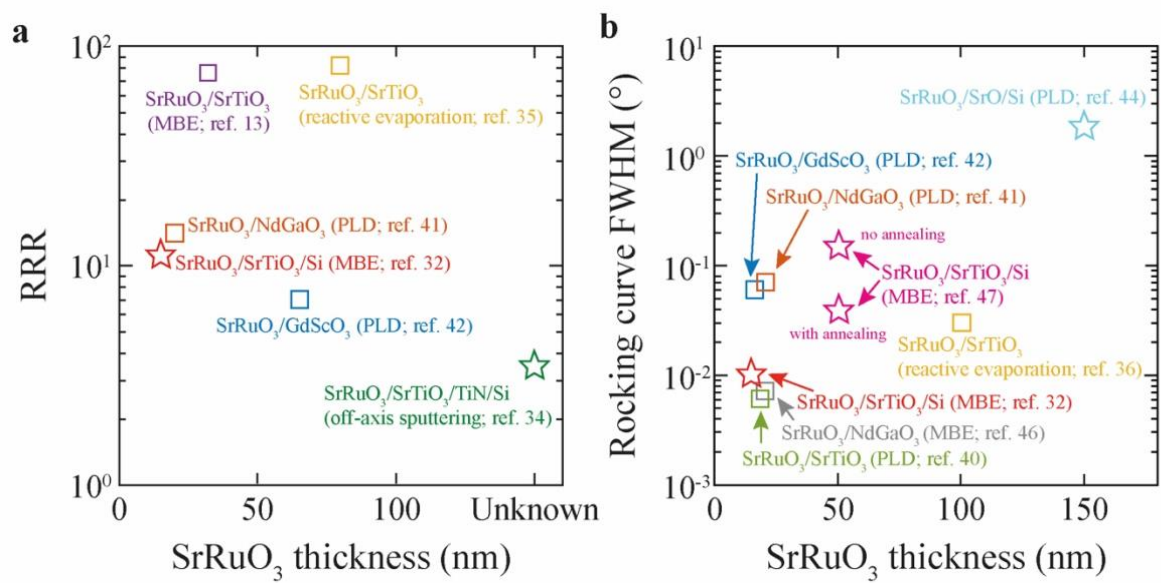

FIG. 2. Transport and structural properties of representative SrRuO$_3$ thin films from the literature. Residual resistivity ratio (a) and full width at half maximum of the rocking curve of the (001)$_{pc}$ or (002)$_{pc}$ peaks (b) for SrRuO$_3$ thin films grown with different techniques and on various substrates as a function of the SrRuO$_3$ thin film thickness. The growth technique and substrate used for each thin film are reported between brackets with the corresponding reference number. [Figure adapted from the Supplementary Material of ref. [32]]. Additional studies to those in panel (a) and reporting SrRuO$_3$ thin films with high RRR (larger than 50) are cited in the main text.

Until recently, most of the attempts done at growing SrRuO$_3$ thin films on Si have resulted in thin films of poor quality both from a structural and an electronic transport point of view. The formation of an amorphous SiO$_2$ layer directly onto Si during the growth of SrRuO$_3$ usually impedes epitaxial growth[32] and results in polycrystalline SrRuO3 thin films with a poor RRR of ~ 3 at most[62]. To achieve epitaxial growth, a multi-step deposition is necessary, where a thin epitaxial buffer layer (e.g., SrTiO3 (001)) is first deposited on Si, which is then followed by the deposition of an epitaxial SrRuO3 thin film onto the buffer layer. This two-step process, however, requires breaking vacuum between the two depositions and hence exposing the surface of the buffer layer to air, which eventually also leads to thin films of poor quality.

Recently, a single-step process has been successfully developed by Wang and co-workers[32], where both the SrTiO3 buffer layer and the SrRuO3 thin films are grown in the same MBE chamber on 2'' commercial Si wafers without breaking vacuum. This approach has resulted in epitaxial SrRuO3 thin films on Si with excellent structural, magnetic and transport properties. Reflection high-energy electron diffraction (RHEED) patterns acquired along the $[100]_{pc}$ and $[110]_{pc}$ azimuths of the SrRuO3 thin films and X-ray diffraction $\phi$ scans demonstrate epitaxial growth of the SrRuO3 thin films with the $[100]_{pc}$ direction of SrRuO3 oriented along the [110] axis of the (001) Si substrate[32]. The RRR of the SrRuO3 thin films reported in this study[32] is of ~ 11 – this is comparable to that of other SrRuO3 thin films grown on single-crystal oxide substrates by PLD[32,40-42] (Fig. 2a). It is worth nothing that MBE is nowadays used also to manufacture semiconductor devices[63], which makes the process reported in ref. [32] appealing for the large-scale production of SrRuO3 devices on Si using the same nanofabrication processes of the CMOS industry.

More recently, machine-learning models have been combined with the MBE technique to quickly determine the growth conditions for high-quality SrRuO3 thin films. Wakabayashi *et al.*, for example, have adopted Bayesian optimization during the MBE growth of SrRuO3 thin films[58]. Their approach consists in applying Bayesian optimization to one growth parameter at a time, whilst keeping all the other growth parameters fixed[58]. Following this procedure, all the MBE growth parameters (e.g., Ru flux rate, growth $T$, and O3-noozle-to-substrate distance) were optimized after only 24 MBE growth runs, and SrRuO3 thin films with a RRR ~ 50 were obtained[58]. Machine-learning-assisted MBE with Bayesian optimization has been reproduced also in other studies[48,49,51,58], and it has yielded SrRuO3 thin films with RRR of ~ 80 and 86 after 35 and 44 MBE optimization runs, respectively[51,64].

It is clear to us that growth optimization of SrRuO3 thin films assisted by machine learning approaches such as Bayesian optimization will eventually replace the typical growth optimization based on a trial-and-error approach. The traditional trial-and-error approach is in fact time consuming and costly, and it ultimately depends on the skills of the researcher carrying out the process.

The ultra-high quality SrRuO$_3$ thin films grown by machine-learning-assisted MBE have also led to the discovery of novel quantum phenomena in SrRuO$_3$. Performing transport measurements on SrRuO$_3$ thin films grown by machine-learning-assisted MBE, Takiguchi and co-workers have shown[51] evidence for Weyl nodes in the electronic band structure of SrRuO$_3$ – the existence of Weyl nodes had only been predicted[65] theoretically in 2013. Weyl nodes are of both fundamental and practical interest because they are tuneable in an applied magnetic field and can provide high-mobility two-dimensional carriers. The two-dimensional nature of these high-mobility carriers stem from Fermi arcs that connect the surface projection of Weyl nodes with opposite chirality. Two recent studies[66,67] have shown evidence for high-mobility two-dimensional carriers from surface Fermi arcs in untwined ultra-high quality SrRuO$_3$ thin films.

Based on the example studies reported above, it is clear that MBE, and in particular machine-learning-assisted MBE, is currently the most reliable technique to produce ultrahigh-quality SrRuO$_3$ thin films. The ultrahigh quality is an essential prerequisite to get access to quantum phenomena recently discovered in SrRuO$_3$ and to develop quantum electronic applications based on transport of Weyl nodes and high-mobility two-dimensional carriers.

MBE, and in particular machine-leaning-assisted MBE, appears therefore as the most promising growth techniques to realize SrRuO$_3$ thin films and devices for quantum electronics. Recent studies[32] have shown that MBE is also suitable for large-scale growth of high-quality SrRuO$_3$ thin films on Si – which is an essential requirement for the integration of SrRuO$_3$ devices with conventional CMOS electronics. In addition to MBE, we believe that other growth techniques are equally promising and should be tested in the future for high-throughput growth of high-quality epitaxial SrRuO$_3$ thin films on Si. These techniques are radiofrequency (RF) magnetron sputtering in a multi-target sputtering chamber equipped with substrate heater and continuous compositional-spread PLD[43] with synchronised translation of the substrate heater with the pulsing of the excimer laser.

**1.2 Structural parameters and experimental tools to control physical properties**

In the previous section, we have described the main physical properties of SrRuO$_3$, the values of the measurable parameters attesting high quality of SrRuO$_3$ thin films, and the growth techniques that can be used to produce such high-quality films. Here, we review the main structural parameters affecting the physical properties of SrRuO$_3$ thin films and we also discuss the experimental tools that can be exploited to control these properties. Achieving fine control over the physical properties of SrRuO$_3$ is in fact another essential ingredient for the development of conventional and quantum electronics applications based on SrRuO$_3$.

As for other perovskite compounds, the physical properties of SrRuO$_3$ depend on a number of structural parameters[2] including the degree of off-stoichiometry, substrate-induced strain, structural disorder, thickness etc. Some properties like the magnetic properties are more sensitive than others in SrRuO$_3$ to any variations in these structural parameters.

Changes in the nominal stoichiometry of SrRuO$_3$ thin films are either due to ruthenium or to oxygen vacancies[45]. The stoichiometry of the SrRuO$_3$ thin films is extremely dependent on the oxygen activity during deposition, which is set by the amounts of atomic and molecular oxygen present during growth[45]. In SrRuO$_3$ thin films made by MBE, nominal stoichiometry is easier to achieve because the fluxes of molecular and atomic oxygen can be controlled independently (also from the Ru and Sr supplied). Atomic oxygen, for example, can be generated in an MBE chamber using a microwave plasma source and its pressure can be tuned by adjusting the oxygen flow supplied to the plasma source and the generator power[45]. At low oxygen activity, in SrRuO$_3$ thin films made by MBE, stoichiometry is mostly set by the amounts of Sr and Ru supplied during growth. At oxygen activities much higher than those suitable for good stoichiometry, Ru vacancies become unavoidable and independent on the amount of Sr and Ru supplied. The increase in Ru vacancies is most likely due to the formation of the volatile compound RuO$_4$, whose concentration increases at higher oxygen activity[45,68].

For SrRuO$_3$ thin films grown by PLD other than MBE, it has been observed that they tend to be normally Ru deficient because a (high) atomic oxygen pressure already exists within the plume, and very little can be done to avoid this[2,45]. This is one of the reasons why, although PLD allows to grow SrRuO$_3$ thin films of consistently good quality, PLD-grown SrRuO$_3$ thin films have normally lower RRR compared to thin films of similar thickness deposited by MBE, where the fluxes of molecular and atomic oxygen can be independently controlled during growth[2,45]. A good crystallinity in SrRuO$_3$ thin films grown by PLD, however, can still be achieved, even in the presence of Ru vacancies[45].

Now we discuss the effect that off-stoichiometry has on the SrRuO$_3$ thin film properties. Ru vacancies induce an expansion in the SrRuO$_3$ unit cell and this is mechanism responsible for a reduction in the $T_{Curie}$, which can be of up to several tens of Kelvin degrees from its bulk value of ~ 160 K (refs. [45,69]). As the amount of Ru vacancies increases, the ratio between the pseudocubic $a_{pc}$ and $c_{pc}$-axis lattice parameters ($c_{pc}/a_{pc}$) becomes lower than 1, and the saturation magnetisation of SrRuO$_3$ increases up to 2.4 $\mu_B$/Ru atom[69]. This suggests that the high spin configuration of half $d$ filled $Ru^4$ ions stabilises[69], as the crystal structure is distorted from $c_{pc}/a_{pc} > 1$ to $c_{pc}/a_{pc} < 1$ by the increase in Ru vacancies.

Oxygen vacancies on their hand cannot be distinguished by Ru vacancies on the basis of lattice parameters[2]. It has been reported, however, that variation in the oxygen stoichiometry of the thin films achieved by varying the oxygen partial pressure, $P(O_2)$, during growth, can influence the $RuO_6$ octahedra rotation and tilting. Like thin films with Ru vacancies, also SrRuO$_3$ thin films with oxygen vacancies exhibit an increase of the $c_{pc}$-axis lattice constant. The increase in the $c_{pc}$-axis lattice constant leads to a deformation of the unit cell from orthorhombic to tetragonal[70,71]. Missing oxygen ions at the octahedral apexes due to oxygen vacancies increase the Ru-Ru repulsion along the $c_{pc}$-axis, which suppresses the rotation of the RuO$_6$ octahedra along the $a_{pc}$- and $b_{pc}$-axes and stabilizes the tetragonal phase[72,73]. These SrRuO$_3$ thin films with a tetragonal structure usually exhibit different electronic transport properties compared to thin films with an orthorhombic unit cell. The different physical properties are also related to differences in the RuO$_6$ octahedra tilting and rotation – which are known to have significant effect on the SrRuO$_3$ properties (as further discussed below).

For thin films grown by PLD, as $P(O_2)$ during growth is reduced, the Sr/Ru ration increases and the structure stabilises into the tetragonal phase[72]. Compared to orthorhombic thin films grown in the same conditions but at higher $P(O_2)$, tetragonal SrRuO$_3$ thin films show an increase in their room-$T$ $\rho$, most likely due to a reduced hybridization between the Ru 4d and O 2p orbitals in the tetragonal phase compared to the orthorhombic one[72]. In addition to the electrical properties, also magnetic properties, and in particular magnetic anisotropy, change as result of the structural phase transition into the tetragonal phase introduced by oxygen vacancies. In a study carried out by W. Lu et al.[70], for example, they show that SrRuO$_3$ thin films with a thickness larger than 50 nm and tetragonal structure have perpendicular magnetic anisotropy, whilst thin films with the same thickness and orthorhombic structure exhibit in-plane magnetic anisotropy[70]. These results also suggest that stochiometric control is a possible route to stabilize the tetragonal phase and the corresponding magnetic anisotropy in SrRuO$_3$ thin films, in addition to varying epitaxial strain or reducing the film thickness[70].

Substrate-induced strain is another parameter that can be tuned to obtain SrRuO$_3$ thin films with desired physical properties for conventional and quantum electronics. Epitaxial strain in stoichiometric SrRuO$_3$ thin films can have a similar effect on magnetism as Ru vacancies in off-stochiometric films, meaning that strain can also induce suppression in $T_{Curie}$ (ref. [34]). The correlation between the structural and physical properties of SrRuO$_3$ with substrate-induced strain has been the subject of several studies[2,34,74-78]. From a structural point of view, there is general agreement that SrRuO$_3$ thin films under substrate-induced tensile strain tend to have a tetragonal structure[41,78], whilst SrRuO$_3$ thin films under substrate-induced compressive strain have an orthorhombic structure[76]. For a fixed growth substrate, strain can also change depending on several growth parameters including the SrRuO$_3$ thin film thickness. SrRuO$_3$ thin films under tensile strain on (110) GdScO$_3$ substrates, for example, show an orthorhombic structure up to a certain thickness (~ 16 nm), beyond which these SrRuO$_3$ thin films assume a tetragonal structure[41].

In general, strain imposed by the substrate, changes the Ru-O and Ru-O-Ru bond lengths, as result of the different rotation of the RuO$_6$ octahedra. In the orthorhombic phase, the RuO$_6$ octahedra rotate out-of-phase about the [010]$_{pc}$ ([1-10]$_{or}$), which is the magnetic easy axis, and in-phase about the [100]$_{pc}$ ([001]$_{or}$) direction, which is the magnetic hard axis[70]. The rotations along these two orthogonal in-plane directions, however, is suppressed in the tetragonal phase[70], meaning that in-plane symmetry breaking is different between the orthorhombic and tetragonal phases. This difference in in-plane symmetry breaking is considered to be the reason for the different magnetic anisotropies observed in orthorhombic and tetragonal SrRuO$_3$ thin films[70,78] (see also above).

Twinning can also have a profound effect on the magnetocrystalline properties of SrRuO$_3$ thin films and introduce anisotropy axes that are different from those of thin films of optimal quality. In the paramagnetic state above $T_{Curie}$, epitaxial SrRuO$_3$ thin films on (001) SrTiO$_3$ which are free of twin-plane defects and with ideal stoichiometry exhibit uniaxial magnetocrystalline anisotropy with an easy axis coinciding with the orthorhombic $b_{or}$-axis[79,80] (i.e., the [010]$_{or}$ axis of the orthorhombic unit cell). We note that epitaxial SrRuO$_3$ thin films grown on (001) SrTiO$_3$ substrates are oriented with the pseudocubic [001]$_{pc}$ axis (equivalent to the orthorhombic [110]$_{or}$ axis) perpendicular to the substrate surface, so that the $b_{or}$-axis is at 45° out of the plane of the film. The uniaxial nature of the magnetocrystalline anisotropy in SrRuO$_3$ has been demonstrated using Lorentz force microscopy[81] as well as through measurements of the magnetic susceptibility χ around $T_{Curie}$. χ shows an increase along the $b_{or}$-axis, as $T$ is decreased from room $T$ down to $T_{Curie}$, by several orders of magnitude compared to its value measured along the $a_{or}$-axis[79]. Below $T_{Curie}$, the easy magnetization axis deviates from the $b_{or}$-axis due to an orientational transition[82] occurring as $T$ is decreased, so that the angle that the easy axis forms with the surface normal, meaning the [001]$_{pc}$- (or [110]$_{or}$-) axis, decreases progressively from ~ 45° to ~ 30° (ref. [10]). Deviations of the angle formed by the magnetic easy axis with the [001]$_{pc}$ ([110]$_{or}$) axis in this films from these values have also been reported, which depend on the presence of intertwined crystal nanodomains[83] or on the crystallographic orientation or on the type of growth mode (e.g., step flow or two dimensional) of the SrRuO$_3$ thin film[80]. In general, changes in the orientation of the magnetic easy axis in SrRuO$_3$ thin films are due to structural deformations of the orthorhombic unit cell[8] (e.g., due to strain). Structural deformations are in turn associated with changes in the rotation and tilt of the RuO$_6$ octahedra[3,84]. As discussed above, when the strain in SrRuO$_3$ thin films changes, the magnetic easy axis of the films can switch from having an out-of-plane to an in-plane orientation[70,84,85].This implies that there exists a strong correlation between spin-orbit interactions in SrRuO$_3$ and its magnetocrystalline anisotropy (ref. [2]).

Variations in the physical properties of SrRuO$_3$ thin films as a function of film thickness have also been intensively investigated. An evolution from a metallic to an insulating behaviour in the electronic transport properties has been observed as the thin film thickness is decreased below a critical value, $d_c$, of a few unit cells (u.c.)[86-88]. The smallest $d_c$ reported to date[89] corresponds to 2 u.c. for bare SrRuO$_3$ thin films, although similar low-$T$ conductivity values to ref.[89] have been obtained for SrRuO$_3$/SrTiO$_3$ superlattices with SrRuO$_3$ thickness of 1 u.c. (ref. [90]). Earlier studies ascribed the thickness dependent of the metal-to-insulator transition (MIT) to several extrinsic mechanisms like disorder, defects, surface electronic reconstruction[86,87,91-92] etc. which become more significant as the SrRuO$_3$ film thickness is reduced and that can lead to an enhancement in weak localization effects[86]. Nonetheless, the atomic-scale precision currently achieved in the growth of ultrathin SrRuO$_3$ thin films rules out extrinsic mechanisms as the origin of the MIT, since a MIT is still observed in ultrathin SrRuO$_3$ grown with state-of-the-art deposition techniques for a thickness below 2 u.c. One of the intrinsic mechanisms that could be responsible for the MIT is the ratio between the Coulomb interaction and the Ru 4d bandwidth resulting from the hybridizations between the Ru 4d and O 2p orbitals[93]. This hybridization is strong and anisotropic in thicker SrRuO$_3$ films generally due to substrate-induced strain, but it becomes weaker in the ultrathin limit leading to Ru-O bonds with an ionic nature and localised Ru 4d orbitals.

The MIT in the electronic transport is also accompanied by a suppression in ferromagnetism, which disappears at a critical thickness below 3 and 4 u.c. in bare SrRuO$_3$ thin films[88,89]. Exchange bias has also been observed in SrRuO$_3$ thin films of thickness smaller than 3 u.c., which points to the possible presence of antiferromagnetic regions in contact with ferromagnetic ones[88]. In a theoretical study[91], it was suggested that ferromagnetism should even persist in thin films of 2 u.c. and that only SrRuO$_3$ thin films with a thickness of 1 u.c. should be non-ferromagnetic due to surface-driven effects. Based on this suggestion, it was recently found[90] that one-unit-cell-thick SrRuO$_3$ films embedded in a SrTiO$_3$/SrRuO$_3$ superlattice, where surface effects are non-existent, are indeed ferromagnetic with a magnetic moment of approximately 0.2 $\mu_B$/Ru atom. This magnetic moment in one-unit-cell-thick SrRuO$_3$ was measured by scanning superconducting quantum interference device (SQUID) microscopy[90].

Magnetic properties in SrRuO$_3$ and possible approaches to tune them have also been the subject of several studies. Despite the numerous studies on ferromagnetism and the recent observation of two-dimensional magnetism[90] in SrRuO$_3$, the itinerant nature of ferromagnetism in SrRuO$_3$ remains not fully understood[94]. Understanding the origin of this magnetism, also studying parent compounds with general structure ARuO$_3$ (A = Sr, Ca, Ba), can prove crucial to access those mechanisms that can be exploited to enhance $T_{Curie}$ in SrRuO$_3$. Enhancing $T_{Curie}$ in SrRuO$_3$ and bringing it closer to room $T$ would significantly extend the number of applications of SrRuO$_3$. Recent studies on parent compounds like CaRuO$_3$ and BaRuO$_3$ have already contributed over the past years to better understand the ferromagnetism in SrRuO$_3$. There exist in fact significant differences between the magnetic properties of all these ARuO$_3$-type compounds. Studying the physical mechanisms behind these differences can help better understand ferromagnetism in SrRuO$_3$ and how to manipulate its $T_{Curie}$.

Although both SrRuO$_3$ and CaRuO$_3$, for example, have an orthorhombic structure with *Pbnm* space group, evidence for ferromagnetic ordering in CaRuO$_3$ has not been found. Several studies[95,96], however, suggest that CaRuO$_3$ is on the verge of ferromagnetic ordering. Unlike SrRuO$_3$, BaRuO$_3$ has a cubic structure with space group *Pm-3m* and it shows ferromagnetic ordering with $T_{Curie}$ ~ 60 K (ref. [97]). The saturation magnetization of BaRuO$_3$ (measured at $T$ = 5 K in an applied $H$ of 5 Tesla) is of ~ 0.8 $\mu_B$/Ru atom[97], which is significantly lower than the saturation magnetization of ~ 1.6 $\mu_B$/Ru atom measured in SrRuO$_3$.

Earlier studies on Sr$_{1-x}$Ca$_x$RuO$_3$ suggested that the suppression of magnetism in Ca$_x$RuO$_3$ is due to a decrease in the Ru-O-Ru bond angle happening for increasing Ca concentration $x$ (the bond angle decreases from 163° in SrRuO$_3$ to 148° in CaRuO$_3$). Theoretical calculations of the band structure of Sr$_{1-x}$Ca$_x$RuO$_3$ compounds also showed that, as the Ru-O-Ru bond angle reduces, the band degeneracy at the Fermi level decreases until the Stoner criterion is no longer verified, and magnetism is suppressed[98]. This explanation, however, cannot account for the reduced $T_{Curie}$ of BaRuO$_3$ compared to SrRuO$_3$ (the Ru-O-Ru bond angle is 180° in BaRuO$_3$), even if size variance effects (induced by Ca and Ba doping) are considered. It should be finally noted that Sr$_{1-y}$Ba$_y$RuO$_3$ compounds show typical Curie-Weiss (CW) behavior for $T > T_{Curie}$ and critical fluctuations[99] near $T_{Curie}$, whereas Sr$_{1-x}$Ca$_x$RuO$_3$ compounds exhibit an unusual $\chi^{-1}(T)$ dependence as $T_{Curie}$ is approached[97].

C. Q. Jin and co-workers[97] have recently argued that the reduction in $T_{Curie}$ for $Sr_{1-y}Ba_yRuO_3$ for increasing Ba concentration $y$ is caused by band broadening induced by Ba doping, since CW behavior persists for $T > T_{Curie}$. This is in contrast with $Sr_{1-x}Ca_xRuO_3$, where Ca doping leads to a reduction in the Ru-O-Ru bond angle and to a dilution of the ferromagnetic interactions, which results in a $\chi^{-1}(T)$ dependence typical of the Griffiths' phase[97]. The arguments proposed by these researchers[97] are also supported by a recent study[100] based on density functional theory + dynamical mean-field theory (DFT+DMFT). These DFT+DMFT calculations suggest that the ferromagnetic transition in $ARuO_3$ ruthenates depends on three parameters: the density of states (DOS) at the Fermi level $E_F$ (in accordance with Stoner's model), the DOS peak position with respect to the ruthenate band edge and its bandwidth. Based on these theoretical models, $CaRuO_3$ has no ferromagnetism due to its large lattice distortion (octahedra tilt and rotation in $CaRuO_3$ is larger than in $SrRuO_3$), which leads to a split of the DOS peak and in turn to a decrease in the DOS at the $E_F$. $BaRuO_3$, which has a large DOS peak, it is also characterized by larger bandwidth and by a DOS peak position further away from the upper band edge than $SrRuO_3$ – these are all factors that result in a suppression of $T_{Curie}$ for $BaRuO_3$ compared to $SrRuO_3$.

In addition to epitaxial strain, which can be used to change the structure and in turn the magnetic and electronic transport properties of $SrRO_3$, reversible control over the same physical properties has also been achieved using an electric field ($E$) applied, for example, via ionic liquid gating (ILG). The earliest studies on the effect of ILG in $SrRuO_3$ already showed that an $E$ applied via ILG can induce large changes in the anomalous Hall component of the transverse resistivity $\rho_{xy}$ (ref. [101]) as well as shifts in the onset $T$ of the MIT and in the $T_{Curie}$ of ultrathin $SrRuO_3$ films[102].

More recently, magneto-ionic effects have been reported by Li and co-workers[103], where an $E$ applied via ILG has been used to move ions (e.g., $H^+$ or $O^{2-}$) in or out of $SrRuO_3$ and induce large changes in the $SrRuO_3$ magnetic state and THEs. In this study[103], the authors have shown that a large $H^+$ gradient induced in $SrRuO_3$ via ILG leads to a protonated compound $H_xSrRuO_3$ with a paramagnetic metallic ground state. The reason for the ferromagnetic-to-paramagnetic phase transition is a change in the electronic band properties induced by a structural change in $SrRuO_3$. As the proton $H^+$ concentration increases under the $V_G$ applied via the ionic liquid, the $c_{pc}$-axis constant in $SrRuO_3$ undergoes an expansion. Theoretical calculations[103] show that, in this distorted structural configuration, the DOS gets strongly modified due to a splitting of the Ru $t_{2g}$ bands which leads to a shift in the spectral weight towards lowers energies. As a result, the DOS changes in such a way that the Stoner criterion for ferromagnetism is not fulfilled, and the paramagnetic ground state becomes energetically favored over the ferromagnetic state.

Also, in ref.[103], at the boundaries of the ferromagnetic-to-paramagnetic phase transition, a hump-like feature is observed in the transverse Hall resistivity $\rho_{xy}$, which Li and co-workers ascribe to a THE. The emergence of a THE is related to an increase in the Dzyaloshinskii-Moriya interaction (DMI) due to inversion symmetry breaking at the ionic liquid/SrRuO$_3$ interface. From this point of view, these results suggests that magneto-ionic effects induced by ILG can be used as an effective tool to reversibly control THEs and the magnetic ground state in SrRuO$_3$ thin films. As for the case of SrRuO$_3$, H$^+$ migration induced by ILG has also been successfully used in the parent compound CaRuO$_3$, where it induces a reversible $E$-driven magnetic transition from the paramagnetic ground state into an exotic ferromagnetic ground state[104].

In addition to reversible $E$-driven variations in H$^+$ concentration under ILG, reversible $E$-driven changes in oxygen vacancies ($V_O$) have also been shown to be an effective tool to vary the DMI strength and to reversibly switch on/off THEs in SrRuO$_3$ heterostructures. To achieve a $E$-tunable modulation in $V_O$, in a recent study[105] a SrRuO$_3$ thin film was grown onto a SrTiO$_3$ substrate, which had been pre-annealed in vacuum to generate a high $V_O$ amount. Since the $V_O$ formation energy in SrRuO$_3$ is lower than in SrTiO$_3$, $V_O$ tend to diffuse from SrTiO$_3$ into SrRuO$_3$ and to accumulate at the SrRuO$_3$/SrTiO$_3$ interface. Under the application of an $E$ ~ 3 kV/cm, in this study[105] J. Lu and co-workers were able to manipulate the $V_O$ concentration at the SrRuO$_3$/SrTiO$_3$ interface and to reversibly enhance or suppress hump-like and bump-like features related to the THE in SrRuO$_3$.

To summarize, in Fig. 3 we show the main physical properties of SrRuO$_3$ and list the structural parameters and experimental tools which can be used to control such properties, as discussed in this section.

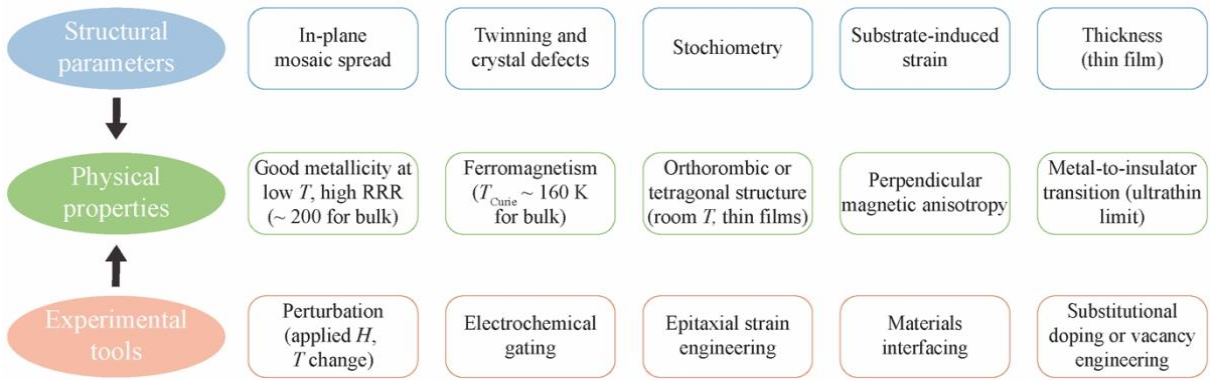

**FIG. 3. Main physical properties of SrRuO$_3$ along with structural parameters and experimental tools that can be used to control them.**

## 1.3 SrRuO$_3$ down to the 0D limit

The physical properties described above together with the mechanisms that can be used to control have been deducted based on extensive studies on 3D SrRuO$_3$ thin films and bulk SrRuO$_3$ single crystals. There are nonetheless other physical properties and effects that emerge in SrRuO$_3$ structures once their dimensionality is reduced. These properties and effects can become particularly relevant when making devices for quantum electronics, which are often based on SrRuO$_3$ systems with dimensionality lower than 3D. In this section, we review the main properties of SrRuO$_3$ that change when reducing its dimensionality from 3D to 0D.

The easiest way to realize two-dimensional (2D) SrRuO$_3$ is by sandwiching a single SrRuO$_3$ layer between two insulating SrTiO$_3$ layers. For this system, it has been theoretically calculated that SrRuO$_3$ should behave as a minority-spin half-metal ferromagnet, with a magnetic moment of $\mu = 2.0$ $\mu_B$/Ru atom[106]. In general, and magnetic reconstruction tend to destroy the metallicity and an insulating behavior is experimentally observed for such 2D SrRuO$_3$ system, albeit with finite low-$T$ conductivity values of ~ 10 $\mu$S (ref. [90]).

Like for other perovskite thin films with general ABO$_3$ structure, the dimensionality of the network formed by the BO$_6$ octahedra (B = Ru for SrRuO$_3$) can also be tuned by growing ABO$_3$/A'B'O$_3$ superlattices and properly varying the orientation of the growth substrate and the periodicity of the superlattice[107]. In the superlattice, the BO$_6$ octahedra normally form a 2D network on a lattice matched (001)-oriented substrate (Fig. 4a), where each octahedra is connected with four others in the *ab*-plane and it is isolated by a B'O$_6$ octahedron along the *c*-axis (i.e., along the growth direction). The 2D network can be reduced to a one-dimensional (1D) network when a (110)-oriented substrate is used, since in this case each BO$_6$ octahedron is only connected to two octahedra along one of the in-plane axes (Fig. 4a). An additional reduction to the 0D regime can be obtained if the superlattice [ABO$_3$]$_1$/[A'B'O$_3$]$_n$ is grown on a (111)-oriented substrate[107] (Fig. 4a). If there are two or more consecutive ABO$_3$ layers with one period, meaning a [ABO$_3$]$_m$/[A'B'O$_3$]$_n$ superlattice (with $m > 1$), then the BO$_6$ octahedra can even be connected in a zig-zag way forming a zig-zag 0D pattern[107] (Fig. 4a).

The reduction in dimensionality of the RuO$_6$ octahedra network leads to a variation in the magnetic properties of SrRuO$_3$ which changes from being a ferromagnetic metal in the 2D limit to an Ising paramagnet in the 1D regime to a ferromagnetic insulator in the 0D case. In the 0D regime, a very significant change in the magnetization has been observed upon strain application[107], which can be exploited in the future for the realization of strain-actuated nanoscale memories[108] based on 0D SrRuO$_3$. Ab-initio calculations also show that half-metallicity and orbital selective quantum confinement can be realized when the dimensionality of $RuO_6$ octahedra network in $SrRuO_3$ is reduced from the 3D to the 0D[107] case.

The 1D growth of $SrRuO_3$ can be also tuned by varying the growth rate and the $SrTiO_3$ (001) substrate miscut angle[109], which in turn determines the height of the 1D steps (Fig. 4b). By further increasing the substrate miscut angle, a bunching of the 1D steps can be obtained[109]. Step bunching in semiconductors and metals has received great attention because bunched surfaces can serve as template for the growth of low-dimensional structures[110,111]. 1D steps of $SrRuO_3$ can therefore be used as template for the epitaxial growth of oxide nanowires including nanowires made of oxide superconductors, which can be investigated for the emergence of topological superconducting phases (see also section 2.4).

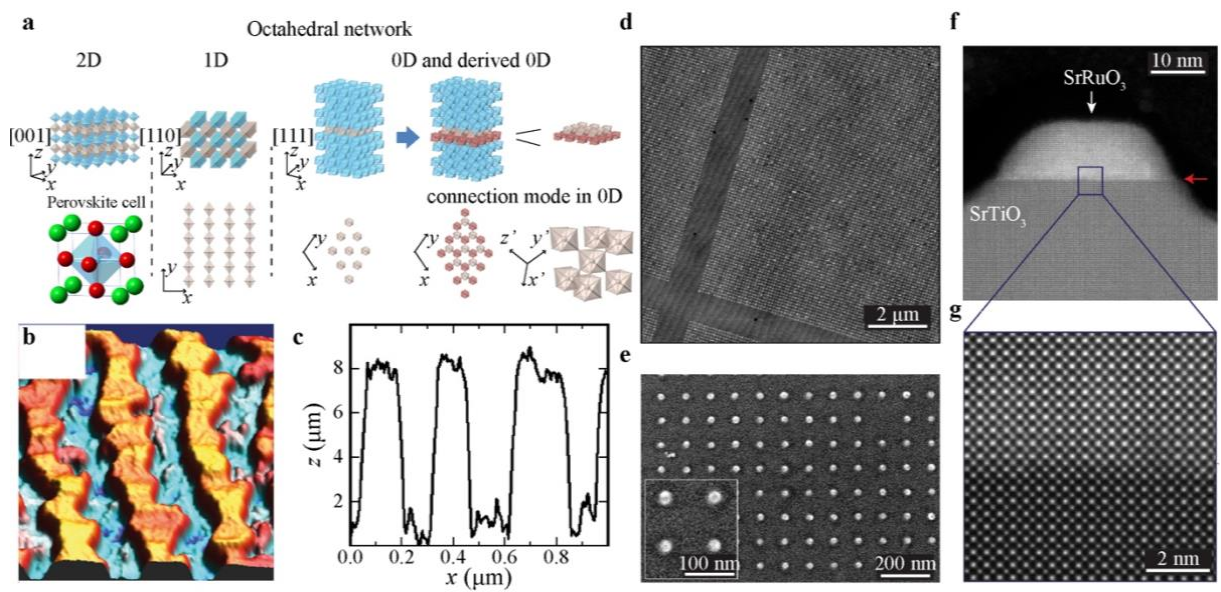

**FIG 4. Realization of $SrRuO_3$ structures with reduced dimensionality. (a) Illustration of different levels of dimensionality obtainable in a $SrRuO_3$-based superlattice for the $RuO_6$ octahedra network [Figure reproduced from ref. [107]]. (b) Atomic force microscopy image showing finger-like 1D $SrRuO_3$ units with height profile (c) measured along a line perpendicular to the one of the finger structures [Figure reproduced from ref. [109]]. Scanning electron microscopy images in (d) and (e) of 0D $SrRuO_3$ nanodots made on a $SrTiO_3$ substrate with high-resolution scanning transmission electron microscopy images of a single nanodot at lower (f) and higher magnification (g). [Figure reproduced from ref. [112]].**

It has also been shown[112] that an array of 0D $SrRuO_3$ nanodots fabricated from a $SrRuO_3$ thin film can exhibit higher $T_{Curie}$ compared to that of the original film (Fig. 4c). The reason for the increase in $T_{Curie}$ is a relaxation of the strain occurring as result of the removal of lateral material around each nanodot[112] compared to the original thin film matrix.

Epitaxial heterostructures of $SrRuO_3/CoFeO_4/BiFeO_3$ have also been used to fabricate nanodots[113] by the nanoporous anodic alumina template method. The array of nanodots shows strong magnetoelectric coupling with clear magnetization switching induced by an applied $E$ – which suggests the possibility of using these 0D SrRuO$_3$ nanodots array for high-density memory storage (> 100 Gbit/in$^2$) or logic devices.

More recently, it has been shown[114] that a single SrRuO$_3$ grain boundary (GB) formed in SrRuO$_3$ grown onto a SrTiO$_3$ bicrystal has transport properties equivalent to that of a spin valve. Apart from highlighting that GBs play a key role towards determining the performance of SrRuO$_3$-based devices, this study[114] suggests that low-dimensionality GBs in SrRuO$_3$ can be used for the realization of novel spintronic devices.

## 2. SrRuO$_3$ in conventional and quantum electronics

After reviewing the physical properties of SrRuO$_3$ and the experimental tools that can be used to control them in section 1, in this section 2 we discuss how SrRuO$_3$ can be combined with other material systems to exploit its physical properties for electronics applications. We do not only illustrate relevant devices that have already been realised, but we also propose devices that have never been made to date. For these new devices, we provide proof-of-concept layouts and explain how they can offer competitive advantages over their equivalents and/or how they can be used in future studies to better understand effects recently discovered in SrRuO$_3$.

Fabrication of devices based on SrRuO$_3$ is nowadays possible thanks to the variety of techniques suitable to make SrRuO$_3$ thin films with excellent properties as well as thanks to the extensive number of studies reported on the optimization of these thin film properties. In addition, the fabrication of SrRuO$_3$ in ultrathin film form and, even more recently, in the form of freestanding oxide nanomembranes have paved the way towards the investigation of material systems, where the reduced dimensionality of SrRuO$_3$ and its interfacing to other oxides have resulted in the discovery of exciting and novel physical effects. The interplay and coexistence within the same material of different types of interactions like spin-orbit interaction, electron-electron correlations, and charge-to-lattice coupling makes SrRuO$_3$ a rich playground for the investigation of a variety of physical phenomena and quantum effects. This wide range of physical phenomena and quantum effects includes orbital magnetic moment and polarization, magnetocrystalline anisotropy, ultranarrow magnetic domains, MIT (as thickness is reduced to the 2D limit), and Berry effects (Fig. 5).

It appears clear to us that achieving control over this rich set of phenomena and effects can lead to the development of devices for conventional electronics (e.g., spin-orbit torque and domain wall spintronics, straintronics) with better performance than existing ones as well as to novel devices for the emerging field of quantum electronics (e.g., topological electronics and superconducting electronics).

In addition to proposing new proof-of-concept electronic devices based on SrRuO$_3$ and to illustrating their layouts, in the following we also describe the materials challenges that have to be addressed to realize such devices. We show that addressing these challenges is crucial to achieve control over the quantum effects and physical phenomena underlying the devices' functioning and ultimately affecting their performance.

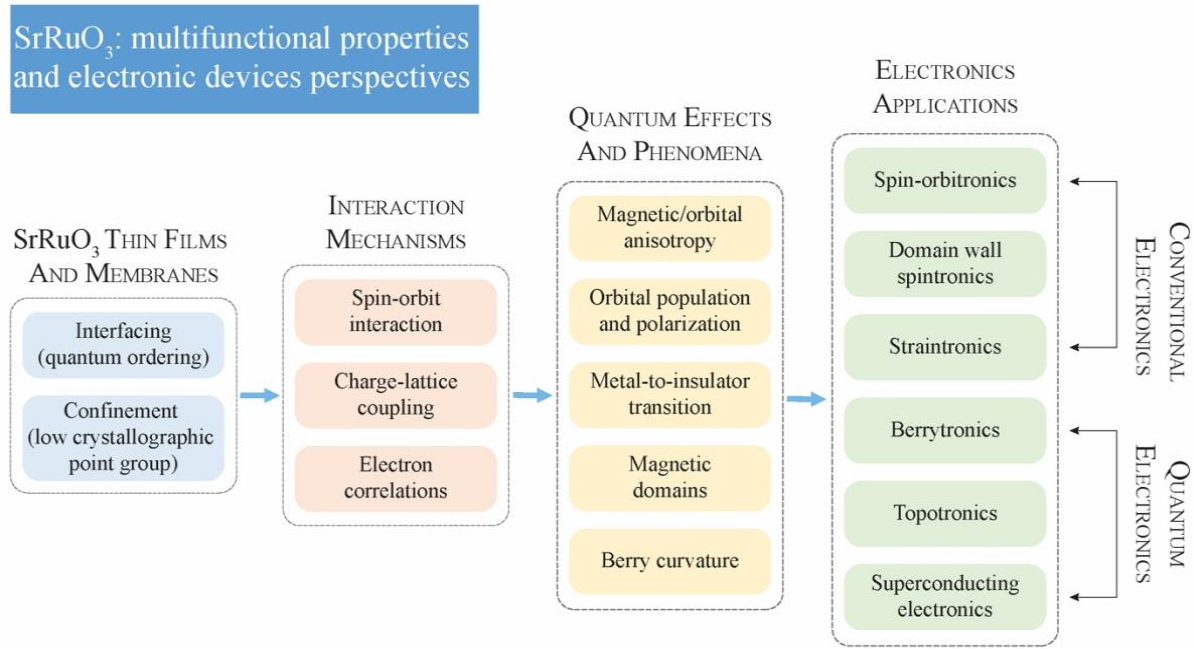

**FIG. 5. Most promising conventional and quantum electronic applications of SrRuO$_3$ (green boxes) based on its physical properties (yellow boxes). The physical properties are determined by the competition of different interaction mechanisms (orange boxes) that can be enhanced and manipulated in ultrathin SrRuO$_3$ structures also through coupling these SrRuO$_3$ structures to other oxide materials (light blue boxes).**

In section 2.1, we suggest new SrRuO$_3$-based devices that are promising for conventional electronic applications, both at room $T$ and cryogenic $T$s. In section 2.2, we illustrate the usage of freestanding SrRuO$_3$ nanomembranes for novel nanoelectromechanical systems (NEMS) with flexomagnetic and flexoelectric properties. In sections 2.3 and 2.4, we propose innovative SrRuO$_3$ devices for quantum electronic applications, which are based on the manipulation of Berry effects in SrRuO$_3$ (section 2.3) or on the coupling of SrRuO$_3$ to superconductors for the generation of topological superconducting states (section 2.4).

## 2.1 Memory and spintronic devices

The application of SrRuO$_3$ for the realization of room-$T$ memory devices is prevented by the low $T_{Curie}$ ~ 160 K of SrRuO$_3$ compared to other 3d-transition metals and ferromagnetic alloys

– which are currently in use for the same applications firstly because they have $T_{Curie}$ higher than room $T$. As a result of this limitation, SrRuO$_3$ has been used mostly as epitaxial metallic electrode for the fabrication of room-$T$ oxide memory devices based on other oxides.

The ever-growing interest in cryogenic electronics, however, is boosting the investigation of energy-efficient and high-density memory technologies that can operate efficiently also at low $T$s. From this point of view and given its high compatibility with other functional oxides like piezoelectrics or ferroelectrics, SrRuO$_3$ can play a major role for the future integration of oxide memory devices in cryogenic CMOS circuits.

We start this section 2.1 by reviewing two applications, one for room-$T$ electronics (i.e., ferroelectric tunnel junctions) and the other for cryogenic electronics (i.e., spin valve devices) where SrRuO$_3$ has been used with good results as metallic electrode and ferromagnetic layer, respectively. Both cases (illustrated in 2.1.1), however, represent examples of applications, where we think that SrRuO$_3$-based devices are unlikely to become de facto technological standards, at least until protocols for large-scale production of SrRuO$_3$ devices with optimal properties and integrable with CMOS technology are developed (as discussed in section 1).

At the end of section 2.1, we outline two other technological applications that may stem from the exploitation of specific SrRuO$_3$ properties. We suggest that two specific properties of SrRuO$_3$, namely its narrow magnetic domains and high spin-orbit coupling, can be used to realise electronic devices that would offer a competitive advantage over existing devices used for the same technological applications.

### 2.1.1. SrRuO$_3$-based devices for conventional electronics with good performance

For memory devices operating at room $T$, one of the applications of SrRuO$_3$ that has already showed good results stems from its use as an epitaxial metallic electrode in ferroelectric tunnel junctions (FTJs). FTJs exploit the change in resistance observed upon polarization reversal of a ferroelectric material to encode digital information. If the ferroelectric is sufficiently thin, by flipping its polarization, it is possible to change its transmission probability for electrons which gives rise to a tunnelling electroresistance effect[115] (Fig. 6a). A variety of FTJs with SrRuO$_3$ used as bottom metallic layer have been already realized, and the list of ferroelectrics used include BaTiO$_3$ (refs. [116-120]), Ba$_x$Sr$_{1-x}$TiO$_3$ (ref. [19]), PbZr$_x$Ti$_{1-x}$O$_3$ (refs. [121-123]), and BiFeO$_3$ (ref. [124]). The changes in resistance observed in these FTJs upon polarization reversal are typically of two orders of magnitude at room $T$ [119], and can increase further at cryogenic $T$s due to a suppression of phonon-assisted indirect tunnelling as $T$ is decreased[120].

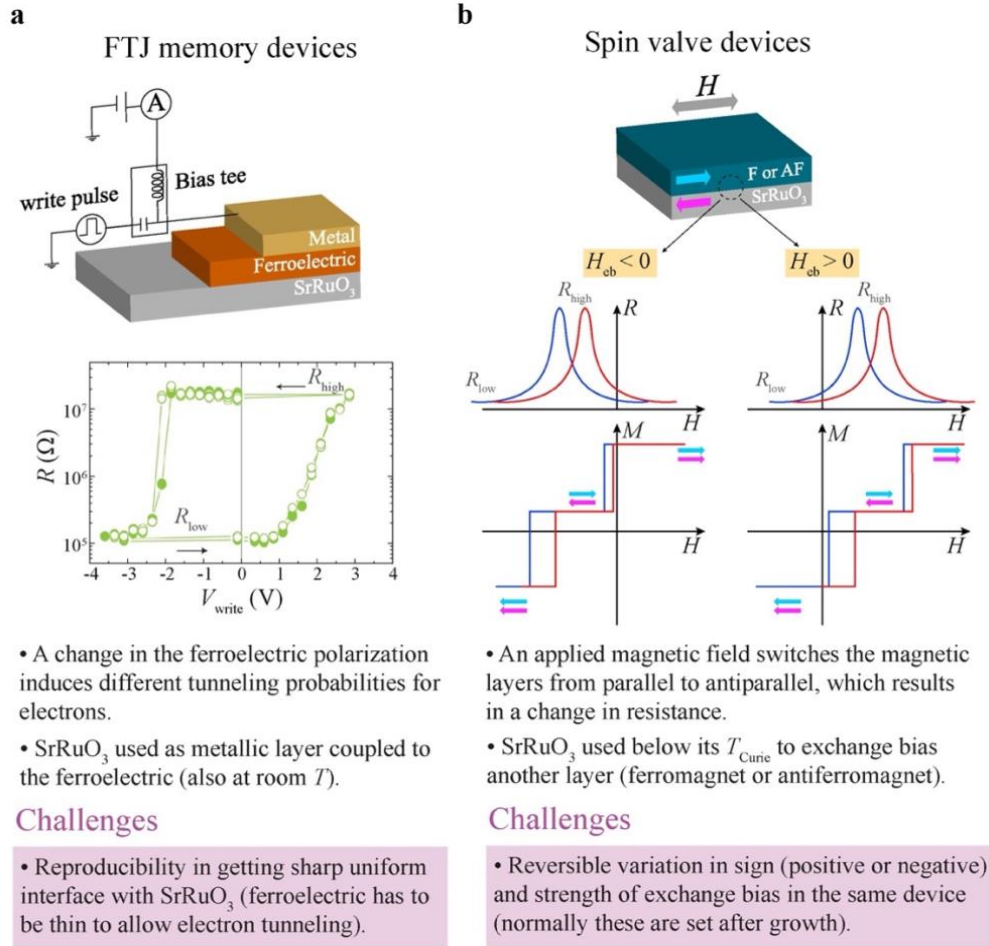

FIG. 6. **Conventional memory and spintronic devices with SrRuO$_3$.** (a) Illustration of a ferroelectric tunnel junction working at room temperature, where SrRuO$_3$ is used as metallic layer for the growth of an oxide ferroelectric. An applied write pulse switches the polarization of the ferroelectric and changes the electron tunneling current between a low- ($R_{low}$) and high-resistance ($R_{high}$) state [Experimental data in (a) reproduced from ref. [115]]. (b) Schematic of a spin valve, where SrRuO$_3$ in its ferromagnetic state is used to exchange bias another ferromagnet (F) or antiferromagnet (AF) layer. Depending on its sign, the exchange bias leads to different asymmetric magnetization versus field, $M(H)$, loops, which in turn define different $H$ ranges for the antiparallel ($R_{high}$) and parallel ($R_{low}$) states of the device.

Although FTJs appear promising for the development of non-volatile resistance-switching random-access memories (RRAMs), both at room $T$ and cryogenic $T$s, the direct current reading of their state is based on the measurement of tunnelling current[124]. This implies that the ferroelectric layer has to be thin to maximise the tunnelling current and facilitate the device readout. Nevertheless, ultrathin ferroelectric barriers exhibit other undesirable effect like a high leakage current which can degrade the device performance[125]. As for other oxide FTJs, also in SrRuO$_3$-based FTJs, ferroelectricity disappears below a critical thickness of the ferroelectric barrier. This critical thickness strongly depends on the uniformity and sharpness of the terminations of the SrRuO$_3$ interface with the ferroelectric – for which oxygen pressure during growth also plays a major role[118]. Achieving fine control over these parameters will play a crucial role towards the development of FTJ RRAMs (at either room $T$ or cryogenic $T$s) based on SrRuO$_3$.

Similar to the case of FTJs, due to its good lattice matching with other oxides, SrRuO$_3$ has also been used below its $T_{Curie}$ to exchange-bias other magnetic oxide thin films (ferromagnets and antiferromagnets) grown epitaxially onto SrRuO$_3$. Exchange bias is typically used in spintronics devices like spin valves to pin the magnetization of a hard ferromagnetic layer, whilst the magnetization of the soft ferromagnetic layer can be switched via an applied magnetic field $H$.

One of the peculiarities of the exchanged-biased heterostructures based on SrRuO$_3$ is that both negative and positive exchange bias can be realised at the interface between SrRuO$_3$ and another magnetic oxide (Fig. 6b). Negative (positive) exchange bias occurs as result of a ferromagnetic (antiferromagnetic) alignment of interfacial spins of the two coupled magnetic materials, and it manifests as a shift of the magnetization hysteresis loop along the same (opposite) direction of the applied cooling $H$. Negative exchange bias has been reported for SrRuO$_3$ epitaxially grown onto the antiferromagnet Sr$_2$YRuO$_6$ (ref.[126]), whilst positive exchange bias has been reported for SrRuO$_3$ grown onto the half-metal ferromagnetic oxide La$_{2/3}$Sr$_{1/3}$MnO$_3$ (refs. [127-130]) or onto Pr$_{0.7}$Ca$_{0.3}$MnO$_3$ (ref. [131]).

In general, positive exchange bias is more difficult to realize experimentally compared to negative exchange bias, and it has been reported only for a few other materials combinations including FeF$_2$/Fe (ref. [132]), Cu$_{1-x}$Mn$_x$/Co (ref. [133]) and Ni$_{81}$Fe$_{19}$/Ir$_{20}$Mn$_{80}$ (ref. [134]) bilayers. The possibility of realising both positive and negative exchange bias when SrRuO$_3$ is coupled to another oxide magnetic material can be used for the realization of novel spin valve devices, where one could switch between different states by varying the sign of the exchange bias.

Like for other coupled ferromagnetic/antiferromagnetic systems, the sign of the exchange bias in SrRuO$_3$-based heterostructures is generally set, once the heterostructure is grown, at given $T$ and cooling $H$ ($H_{cool}$). Under fixed $H_{cool}$ and $T$, the sign of the exchange bias in SrRuO$_3$ heterostructures is set by interfacial mixing and layer thickness which are fixed after growth – these parameters affecting magnetocrystalline anisotropy and layer magnetization[128,129,131]. Within the same heterostructure, however, the sign of the exchange bias can be changed upon varying $H_{cool}$ or $T$. In general, for a bilayer system consisting of two coupled magnetic materials, a large enough $H_{cool}$ can induce either a negative exchange bias for a ferromagnetic-like interface coupling, or positive exchange bias for antiferromagnetic-like interface coupling.

In a few systems, a sign change in the exchange bias has been observed upon increasing $H_{cool}$ – which is typically due to the formation of domain walls parallel or antiparallel to the bilayer interface[135]. For SrRuO$_3$-based systems, a change in the sign of the exchange bias induced by a variation in $H_{cool}$ has been reported, for example, in SrRuO$_3$/PrMnO$_3$ superlattices[136]. Here, the antiferromagnetic exchange coupling switches from positive to negative upon increasing $H_{cool}$. Controlling the sign of the exchange bias by varying $H_{cool}$, however, is not practical for applications. This is because such approach usually involves increasing the operational $T$ of the device about $T_{Curie}$, changing $H_{cool}$ and cooling down the device again. Reversible control of the exchange bias sign (e.g., voltage-driven) at a fixed operational $T$ would be better for technological applications.

A reversible voltage-driven switching in the sign of the exchange bias has been recently achieved[137] at room $T$ in systems not including SrRuO$_3$ and consisting of the antiferromagnet BiFeO$_3$ coupled to ferromagnets like Co$_{0.9}$Fe$_{0.1}$ or La$_{2/3}$Sr$_{1/3}$MnO$_3$. This type of devices, which rely on the strong coupling existing in the multiferroic BiFeO$_3$ between ferroelectric and antiferromagnetic order, can be operated at room $T$ and pave the way for a new type of ultralow energy non-volatile memory. It seems very unlikely tto us hat similar devices based on SrRuO$_3$ can be developed and gain a competitive advantage over devices like those in ref. [137].

## 2.1.2. SrRuO$_3$-based devices for conventional electronics with competitive advantage

The two applications discussed in section 2.1.1 and shown in Fig. 6 are less likely to be carried out with SrRuO$_3$-based devices other than with already existing devices based on other materials. Nonetheless, we identify two other applications for conventional electronics, where SrRuO$_3$ devices can offer better performance than existing devices and become the better alternative, once high reproducibility and scalability in their fabrication is also achieved.

The first application which we illustrate stems from a characteristic magnetic property of SrRuO$_3$, consisting in its domain walls being much narrower than in other oxide ferromagnets. The narrow domain walls of SrRuO$_3$ can be used for low-dissipation racetrack cryogenic memories. In racetrack memories based on domain wall motion, data bits are stored in the form of magnetic domains that are then moved along a nanowire strip, typically through the application of a current[138]. The main issue of the racetrack memories proposed to date, however, is the significant Joule heating induced by the large currents which are typically required to move the magnetic domains through small nanowires[139]. Thanks to its small domain wall width (DWW), SrRuO$_3$ can be potentially used to overcome this issue and realise racetrack memories with lower energy dissipation than those proposed to date. It has been already reported, for example, that domain wall motion in SrRuO$_3$ can be induced with a current density that is at least one order of magnitude lower than that needed for ferromagnetic metals with similar depinning fields[140,141]. Although an exact measurement of the DWW in SrRuO$_3$ has still not been done, it has been estimated[140] that the DWW in SrRuO$_3$ can be as low as 3 nm at $T < 100$ K. An upper limit of 10 nm has also been estimated for DWW in SrRuO$_3$ based on scanning tunnelling spectroscopy (STS) measurements[142] at 4.2 K on SrRuO$_3$/YBa$_2$Cu$_3$O$_7$ bilayers. The upper limit for DWW in this STS study[142] has been determined by studying the spatial variation of the superconducting gap induced in SrRuO$_3$ by YBa$_2$Cu$_3$O$_7$ (YBCO) via the proximity effect. A more accurate measurement of the DWW in SrRuO$_3$ can be carried out nowadays using local magnetometry techniques with very high spatial resolution like nitrogen vacancy magnetometry[143].

The second use case of SrRuO$_3$ for conventional electronics which we identify is the realization of spin-orbit torque (SOT) spintronic devices based on SrRuO$_3$ with electrical control of their state. To date, SOT devices have been fabricated mostly with heavy metals (e.g., Pt, Ta, W, Bi etc.) or semiconductors, whilst oxides have remained mostly unexplored for SOT spintronics. Oxide materials like SrRuO$_3$ or SrIrO$_3$, however, also exhibit a large spin Hall conductivity, $\sigma_{SH}$, and can be therefore used as source of SOT. Spin-torque ferromagnetic resonance (ST-FMR) experiments[144] done on Co/SrRuO$_3$ show that SrRuO$_3$ has a $\sigma_{SH} > 1.5 \times 10^3$ $\hbar/e$ S cm$^{-1}$ ($\hbar = 1.05 \times 10^{-34}$ J s being the reduced Planck constant) with a spin-Hall angle $\theta_{SH} > 0.24$ at $T = 60$ K – these are values comparable to those of heavy metals.

In addition to the large $\sigma_{SH}$, other studies also show there exists a direct correlation in SrRuO$_3$ between the spin orbit interaction and the rotation and tilting of the RuO$_6$ octahedra[144]. To achieve precise tunability over the SOT strength, further investigation is required, since it is currently difficult to disentangle all the mechanisms affecting the SOT strength (e.g., strain, RuO$_6$ octahedra rotations).

Independently on the physical mechanisms (or combinations thereof) affecting SOT, it seems that the SOT strength can be tuned electrically. Some mechanisms like the RuO$_6$ octahedra rotation and strain, which seem to affect the SOT strength, can be indeed controlled electrically. One possible way to achieve the electrical control is by applying a voltage to a piezoelectric exerting strain onto SrRuO$_3$, as sketched in Figs. 7d and 7e. We think that achieving electrical tunability of the SOT strength in SrRuO$_3$ will pave the way for new SOT spintronic devices with electrical control of their state.

By carefully engineering the SrRuO$_3$ strain in Ni$_{81}$Fe$_{19}$/SrRuO$_3$ bilayers and using a combination of ST-FMR and in-plane harmonic Hall voltage measurements[146,147], it has already been shown that the SOT efficiency and $\sigma_{SH}$ can increase by almost two orders of magnitudes. The authors of these studies[146,147] correlate the increase in SOT strength and $\sigma_{SH}$ with a change in the crystal structure of SrRuO$_3$ from orthorhombic (under compressive strain) to tetragonal (under tensile strain). We note that these large $\sigma_{SH}$ (up to ~ 441 × $\hbar/e$ S cm$^{-1}$) and SOT efficiency (up to ~ 0.89) values [146,147] have been measured for SrRuO$_3$ at room $T$.

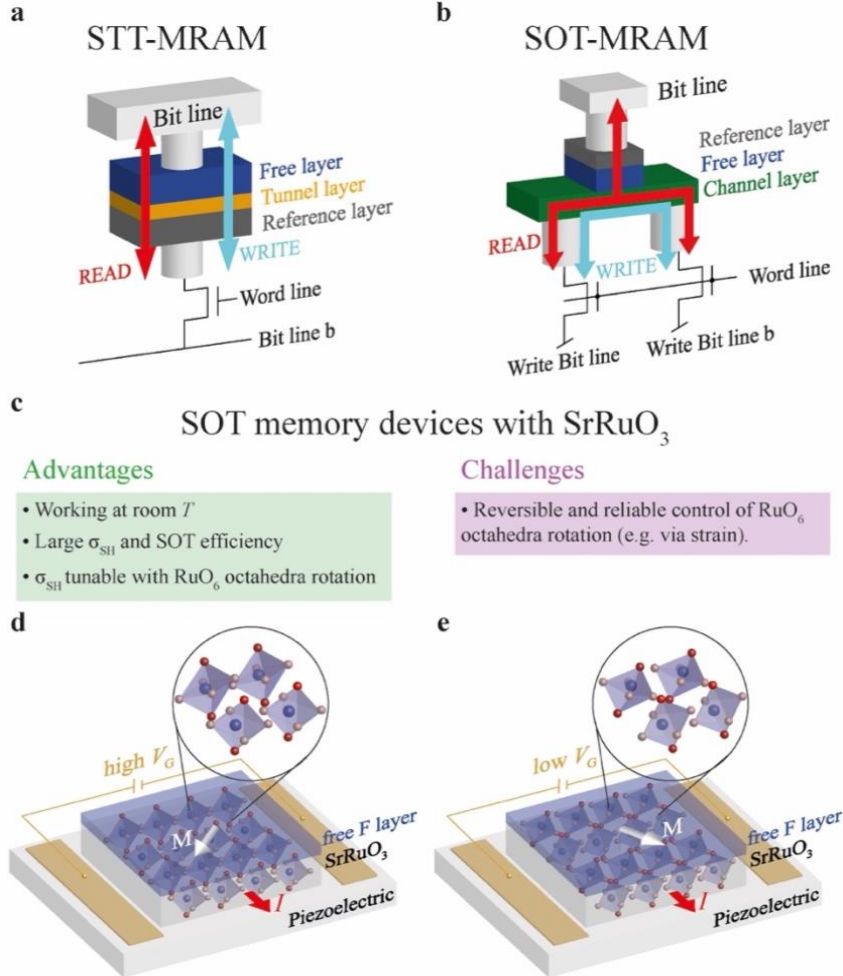

FIG. 7. Application of SrRuO$_3$ for spin-orbit torque memories. Illustration of a typical spin-transfer torque magnetoresistive random access memory (STT-MRAM) device in (a) and of a spin-orbit torque- (SOT-) MRAM device in (b). [Panels (a) and (b) are adapted from ref. [148]]. In a STT-MRAM device the switching of the magnetization of the free magnetic layer is obtained via a tunnelling current injected through a magnetic tunnel junction, whereas in a SOT-MRAM device the current is injected through a layer with high spin orbit coupling that exerts SOT on the free layer. (c) Main advantages and challenges for the realization of SOT-MRAM devices based on SrRuO$_3$ with the layout shown in (d) and (e), where a gate voltage $V_G$ applied to the piezoelectric is used to reversible switch the RuO$_6$ octahedra and tilt them between two configurations. Each configuration leads to a different SOT on the free magnetic layer, allowing to electrically switch the device between two states.

Based on the above results and considerations, we envision that the SrRuO$_3$-based devices that we propose with high SOT efficiency can be used also at room $T$. As a result, these devices can find application in the next generation of SOT- magnetoresistive random-access memories (SOT-MRAMs). SOT-MRAM been recently proposed to overcome the major limitations of spin-transfer torque memories (STT-MRAMs)[148], which represents the current state-of-the-art MRAM technology[149] (Fig. 7a). STT-MRAM has already entered volume production in all major foundries, also thanks to its compatibility with CMOS technology[149].

The main limitations of STT-MRAM are related to large switching currents needed for STT-MRAM operation. These large switching current prevent the application of STT-MRAM for ultra-fast operations at the sub-nanosecond regime, also due to the stochastic nature of STT. In addition, large switching currents also generate reliability issues because they have to flow through the thin oxide of the magnetic tunnel junction (MTJ) as shown in Fig. 7a, which reduces the MRAM endurance over time. By contrast, the switching current in SOT-MRAM does not flow across the MTJ, but through a heavy metal or another material coupled to the magnetic free layer (Fig. 7b).

To make the switching of the free layer magnetization in a SOT-MRAM device more deterministic, a small $H$ field is often applied perpendicular to the free layer. Several $H$-free schemes have also been proposed[150-153], but these usually result in a more complex memory cell fabrication. Recently, a $H$-free switching of the perpendicular magnetization in $SrRuO_3$ was achieved in $WTe_2/SrRuO_3$ bilayers at 40 K, where the $WTe_2$ acts as source of out-of-plane spin polarization due to its reduced crystal symmetry[154].

We envision new SOT devices where the SOT strength in $SrRuO_3$ can be tuned, also at room $T$, via voltage-driven strain exerted by a piezoelectric coupled to $SrRuO_3$ (Figs. 7d and e). The voltage-driven modulation in SOT strength leads to a change of the switching current for the magnetization of a free layer grown onto $SrRuO_3$ between two values. The bistability in the switching current is used to reversibly switch the SOT device between two states.

## 2.2 Straintronics

The possibility of modulating the spin-orbit interaction in $SrRuO_3$ by inducing structural distortions in the material can also be exploited for the realization of novel transducers, actuators, and sensors. Shape memory effect materials like Heusler compounds, which exhibit changes in their shape in response to the application of an external stimulus (e.g., temperature, magnetic field, strain) are nowadays already studied for these applications.

Although shape memory effects are rare in oxides with the only exception of oxide multiferroics, they have recently been observed[155] also in $SrRuO_3$. In $SrRuO_3$, shape memory effects emerge possibly due to a combination of the strong spin-orbit interaction with a weak pinning of the magnetic domain walls. It has been shown that, upon field cooling SrRuO$_3$ in a $H$ of ~ 1 Tesla applied along the [110]$_{pc}$ axis, a single domain state can be induced in SrRuO$_3$, as result of the growth of domains parallel to the applied $H$[155]. Unlike for Heusler alloys[156,157], SrRuO$_3$ remains in this structurally distorted phase, which is stable at low $T$s and against magnetic field sweeps. Upon warming above $T_{Curie}$[155], SrRuO$_3$ exhibits a shape memory effect and relaxes back from a single domain into a multidomain state configuration.

It has been recently shown that epitaxial strain can also be used as an effective tool to vary the magnitude and sign of the Berry curvature and in turn modulate related effects. Several groups had already demonstrated that epitaxial strain affects the magnetic properties of SrRuO$_3$ thin films[41,74,78,158-161]. The magnetic properties are affected by strain due to the strong coupling existing between lattice distortion and electronic band structure in SrRuO$_3$. In their recent study, Wakabayashi and co-workers[53] have performed a systematic investigation of the effect of epitaxial strain on the electrical and magnetic properties of ultrahigh-quality SrRuO$_3$ thin films. These thin films were deposited using machine-learning-assisted MBE on various perovskite substrates with mismatch ranging from -1.6% to 2.3% (compared to bulk SrRuO$_3$). Following this approach, the authors could single out all the effects that strain induces on its own on magnetic and transport properties in SrRuO$_3$. All the other concurrent factors typically affecting magnetic and transport properties (e.g., defects, off-stoichiometry etc.) were in fact not present in these thin films due to their ultrahigh quality.

Motivated by these previous results and by the fact that Berry effects are also very sensitive to changes in the electronic band structure, Tian and co-workers[162] have recently investigated the effect of epitaxial strain on the AHE in both tensile- and compressive-strained SrRuO$_3$. In their study[162], they have found that epitaxial strain can be used as a tool to manipulate the Berry curvature, and the corresponding AHE (in amplitude and sign). Consistently with previous reports[41,78], Tian and co-workers have shown that, as the strain changes from compressive to tensile, the magnetic easy axis of the SrRuO$_3$ thin films changes from an out-of-plane to an in-plane orientation. The change in magnetic anisotropy is accompanied by a variation in the sign and amplitude of the anomalous Hall component of the transverse Hall resistivity $\rho_{xy}$. Ab-initio calculations suggest that the change in $\rho_{xy}$ is due to change in the energies of Ru d-orbitals, whose rotation varies under strain. The rotation of the Ru moments in real space in SrRuO$_3$ thin films under tensile strain affects their Berry curvature. For thin films under tensile strain, the $\rho_{xy}$ versus $H$, $\rho_{xy}(H)$, curves exhibit a non-monotonic trend in the intermediate $H$ region at low $T$s, in contrast with the typical hysteretic behavior expected for an AHE, which recovers in $\rho_{xy}(H)$ for SrRuO$_3$ thin films under compressive strain. Also, whilst for compressive-strained SrRuO$_3$ thin films, ρ$_{xy}$ changes sign with $T$ and it goes from positive to negative at a $T$ typically of ~ 125 K before becoming null at $T_{Curie}$, for tensile-strained thin films ρ$_{xy}$ is negative independently on $T$. These results reported by Tian et al.[162] pave the way for the application of epitaxial strain engineering to reversibly control AHEs in SrRuO$_3$-based devices.

By applying strain to SrRuO$_3$ in the form of freestanding nanomembrane, it should be possible to achieve larger variations in the SrRuO$_3$ crystallographic structure, and in turn a larger modulation of the SrRuO$_3$ physical properties (transport and magnetic) and of related effects (e.g., AHEs and THEs). Free-standing single-crystal oxide membranes of various materials including SrRuO$_3$ have already been fabricated either via either chemical or mechanical lift-off[163-169] from the growth substrate (Fig. 8). Both processes are non-destructive, unlike other physical release methods used for silicon-on-insulator technology[170] or for light-emitting diodes based on GaN[171]. Freestanding oxide nanomembranes can be made without any thickness limitations down to the monolayer limit[167] and can sustain strain up to 8% (ref.[169]), which is unachievable through conventional strain engineering of thin-film heteroepitaxy. In addition to the large strain that can be exerted onto them, SrRuO$_3$ oxide nanomembranes can be stacked onto materials that are difficult to grow epitaxially onto SrRuO$_3$ either because they have different lattice parameters or because they are stable under different growth conditions[163,164].

The fabrication of freestanding SrRuO$_3$ nanomembranes using the chemical lift-off approach has already been reported by several groups[168,172-176]. In all these cases, the SrRuO$_3$ thin film has been grown onto a lattice-matched sacrificial layer which is grown, without breaking vacuum, in between the substrate and SrRuO$_3$. To date, the sacrificial layer that has been mostly used is Sr$_3$Al$_2$O$_6$, which can be dissolved in water as illustrated in Fig. 8a. Nonetheless, the water solubility of Sr$_3$Al$_2$O$_6$ also represents a limiting factor for practical applications due to its instability in air. A more stable sacrificial layer is the brownmillerite SrCoO$_{2.5}$, which has been successfully used by H. Peng and co-workers[176] for the fabrication of freestanding SrRuO$_3$ nanomembranes with (001)$_{pc}$, (110)$_{pc}$ and (111)$_{pc}$ orientations. Unlike Sr$_3$Al$_2$O$_6$, SrCoO$_{2.5}$ is stable in air[176], which is important from an application-related perspective. Also, SrCoO$_{2.5}$ can be dissolved using eco-friendly weak acid solutions such as vinegar, carbonated drinks and acetic acid[176].

Compared to the epitaxial SrRuO$_3$ thin films before lift-off, free-standing SrRuO$_3$ nanomembranes may exhibit a lower spin state with a moment of 1.0 μ$_B$/Ru atom and 1.7 μ$_B$/Ru atom for the (011)$_{pc}$-oriented and (111)$_{pc}$-oriented nanomembranes, respectively[174]. (001)$_{pc}$-oriented SrRuO$_3$ nanomembranes fabricated by chemical lift off with a SrCoO$_{2.5}$ sacrificial layer exhibit remarkable changes in magnetic anisotropy compared to the SrRuO$_3$ thin films before lift off[176]. Changes in magnetic anisotropy, however, are not observed for (110)$_{pc}$- and (111)$_{pc}$-oriented SrRuO$_3$ nanomembranes made following the same chemical lift-off procedure[176].

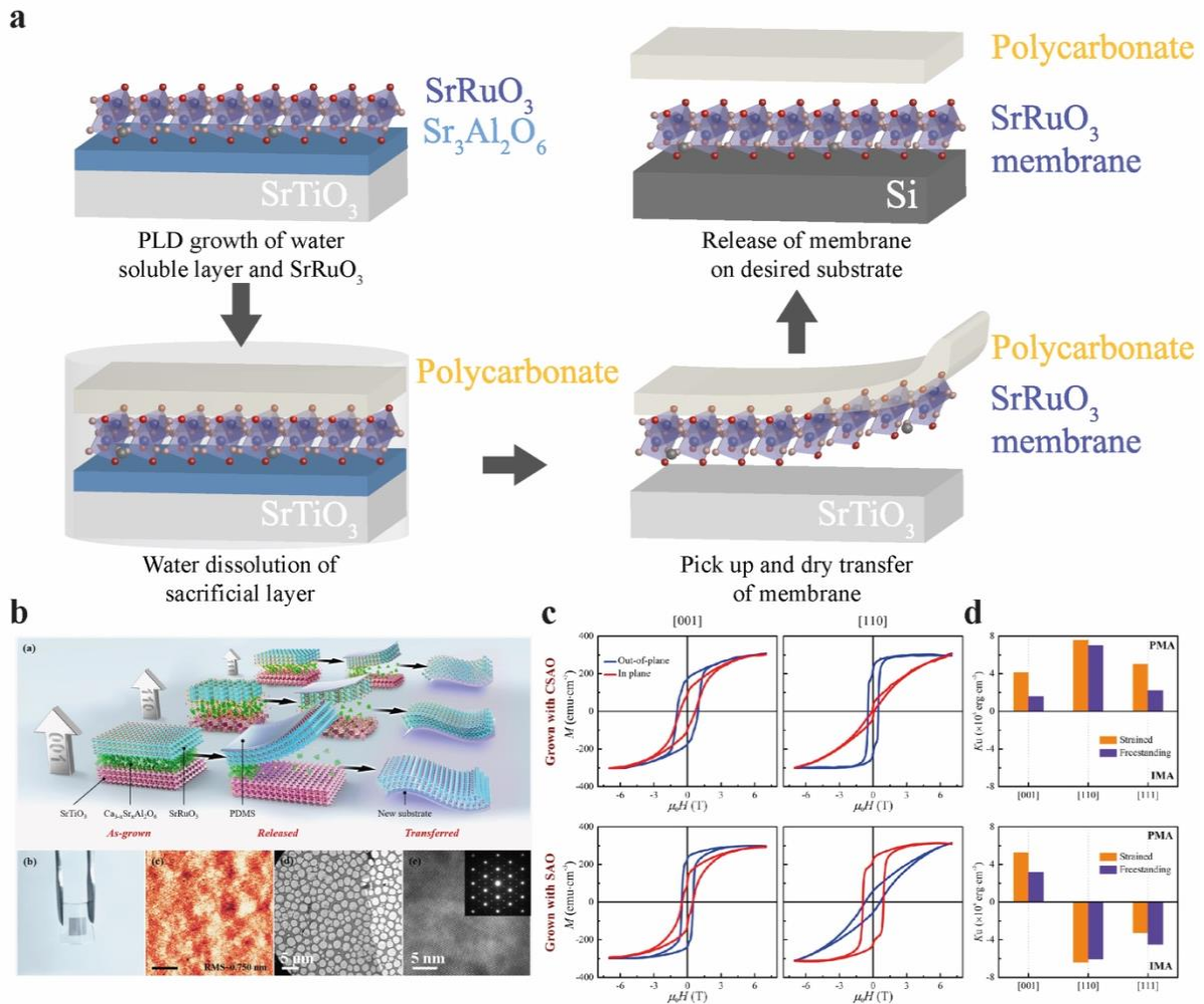

FIG. 8. Fabrication of free-standing single-crystal SrRuO$_3$ nanomembranes. (a) Steps for the fabrication of a SrRuO$_3$ nanomembrane consisting of deposition of a SrRuO$_3$ thin film on a water-soluble thin film (e.g., Sr$_3$Al$_2$O$_6$) epitaxially grown on a SrTiO$_3$ substrate, dissolution of the sacrificial layer in water, pick up of the nanomembrane with polycarbonate and dry transfer of the nanomembrane onto the desired substrate (e.g., Si). (b) Illustration of fabrication of SrRuO$_3$ nanomembranes with different orientations, optical image of a nanomembrane on polydimethylsiloxane (PDMS), characterization of the structural topography of the nanomembrane via atomic force microscopy, scanning electron microscopy and transmission electron microscopy. (c) Magnetization versus applied field loops for SrRuO$_3$ nanomembranes with different orientations and grown onto different sacrificial layers, and comparison of their in-plane and perpendicular magnetic anisotropy with those of the original SrRuO$_3$ thin films before dissolution of the sacrificial layer in water in (d). [Panels from (b) to (d) adapted from ref. [168]].

Probing the $T$ evolution of the mechanical response of resonators fabricated from freestanding SrRuO$_3$ nanomembranes through laser interferometry, it has been also shown that structural phase transitions occurring in SrRuO$_3$ can be identified[175]. This approach suggests a novel method to identify phase transition occurring in complex oxide systems based on freestanding nanomembranes.

SrRuO$_3$ nanomembranes and related heterostructures can be applied for high-performance NEMS including suspended Bragg reflectors, piezoelectric sensors, and mechanical transducers. These NEMS devices can be also integrated with CMOS devices, if the epitaxial growth of SrRuO$_3$ at high $T$ is carried out separately from the rest of the CMOS fabrication processes.

Although SrRuO$_3$ nanomembranes have not yet been tested for NEMS applications, it has already been shown that nanomembranes of other oxides such as SrTiO$_3$, which like SrRuO$_3$ is neither piezoelectric nor ferroelectric, have a flexoelectric figure of merit (i.e., the curvature of the nanomembrane divided by the $E$). The figure of merit of these SrTiO$_3$ nanomembranes are comparable to those of the best piezoelectric and ferroelectric NEMS reported in the literature[164]. Since the figure of merit of a NEMS scales up with the inverse of the cube of the thickness[164] and the oxide nanomembranes tested to date for NEMS are of ~ 100 nm in thickness[177], SrRuO$_3$ nanomembranes of a few tens of nanometres in thickness can provide a much higher figure of merit compared to existing state-of-the-art flexoelectric NEMS.

Freestanding of ultrathin SrRuO$_3$ can also be stacked onto ultrathin nanomembranes of other oxide materials, as it is done to make heterostructures of 2D van-der-Waals (vdW) materials. Heteroepitaxial oxide nanomembranes with SrRuO$_3$ can be tested, apart from their flexoelectric figure of merit, also for their flexomagnetic properties, meaning for an increase in their magnetization under an applied strain gradient (Fig. 9). Flexomagnetism has not been yet estimated nor observed in complex oxides, but SrRuO$_3$ may exhibit large flexomagnetic effects due to its strong coupling of lattice and spin degrees of freedom. Flexomagnetic SrRuO$_3$-based NEMS devices can be potentially used for the realization of magnetic sensors with extremely high sensitivity[178] and resonant frequency tuneable over a very wide frequency range[179].

## 2.3 Berrytronics

Engineering non-collinear magnetic textures and achieving control over topological effects correlated to them has emerged as a promising route for the development of novel quantum electronic devices. Studies triggered by these motivations have also led to the discovery of new phases in condensed matter, which is crucial for the development of quantum technologies.

In the ongoing studies on topology associated to non-collinear spin textures, SrRuO$_3$ has gained a primary role. SrRuO$_3$-based heterostructures with strong inversion symmetry breaking and spin-orbit coupling can be engineered[180-182]. Strong inversion symmetry breaking, and spin-orbit coupling are key ingredients to generate spin textures that are non-collinear in real space and have a topological character.

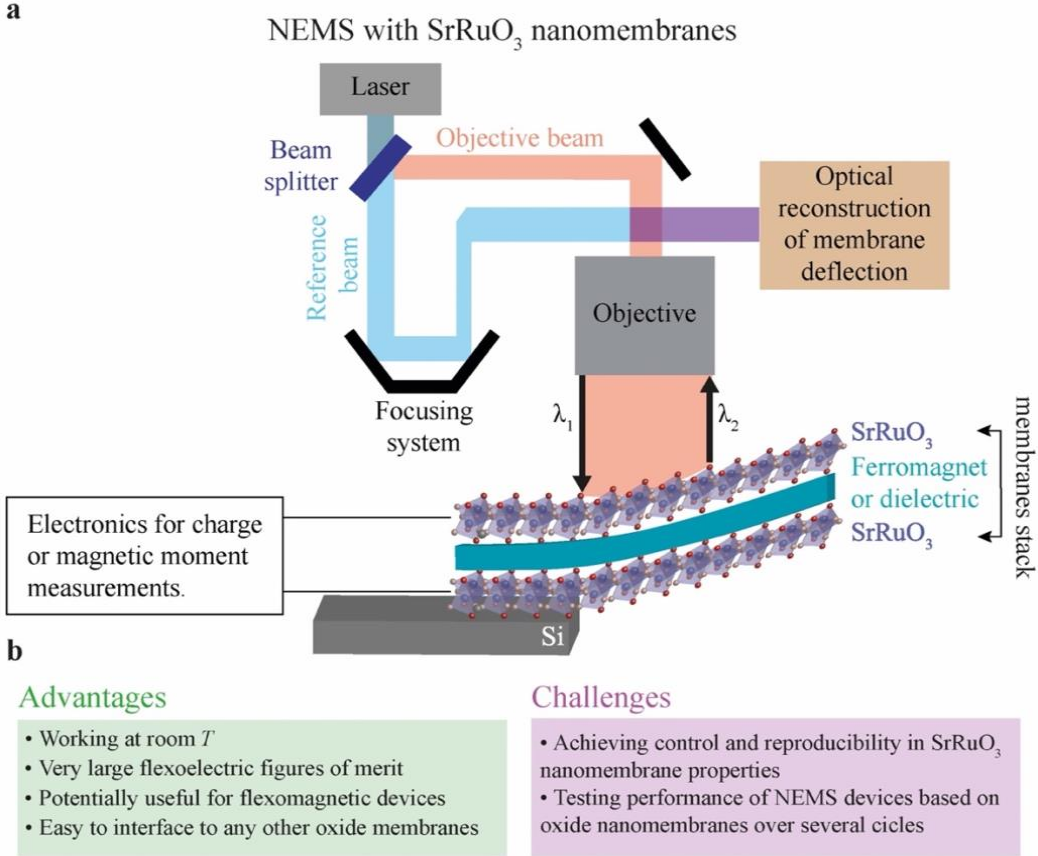

**FIG. 9. Realization and testing of NEMS devices with SrRuO₃ oxide nanomembranes. (a) Illustration of a setup to measure the bending of a NEMS device based on a SrRuO₃ oxide nanomembrane heterostructure: the objective beam is focused onto the sample and the light reflected is collected to form an interference pattern with the reference beam. Any height difference in the nanomembrane stack induces a phase difference in the light reflected. The nanomembrane device is connected to electronics for the measurement of charge or magnetic moment variations of the oxide nanomembrane (dielectric or ferromagnet) sandwiched between metallic SrRuO₃ nanomembranes [Figure adapted from ref.[177]]. The SrRuO₃ nanomembrane can also be used on its own (i.e., without stack) and tested for its flexoelectric (at room $T$ or below) and flexomagnetic properties (below $T_{Curie}$). (b) Main advantages and challenges of NEMS devices made from SrRuO₃-based oxide nanomembrane heterostructures.**

Most of the topological and spin-transport phenomena studied in SrRuO₃ are intimately related to the curvature of a band structure property of materials known as Berry phase ($\Phi_B$) and to its curvature $\Omega_B$, which in SrRuO₃ is non-null. $\Phi_B$ is a geometric quantum phase[183], while $\Omega_B$ (Fig. 10a) acts for electrons as the equivalent of a fictitious magnetic field that introduces a transverse velocity component to the electrons' motion. A non-null curvature of $\Phi_B$ can contribute to electronic quantum effects like the AHE[184], the Spin Hall Effect[185-187] (SHE), the THE[188-189], the Quantum Hall Effect[190] (QHE) and affect a variety of other phenomena[183] like orbital magnetism[191-192], electrical polarization[193-194], quantum charge pumping[195], and topological superconducting phases[196-197].

The $\Omega_B$ in momentum- ($k$-) space induces a cyclotron motion of electronic modes around a crossing point that gives a nonzero intercept in the Landau level phase diagram. The existence of this motion has been verified experimentally in transport experiments through measurements of Shubnikov-de Haas oscillations[181,198]. Sources of either time-reversal or inversion symmetry breaking can lead to a non-zero $\Omega_B$ that is locally distributed in the Brillouin zone and can manifest with a strong spin-orbital and band dependence. Regardless of the magnetization of a material, a modulation in $\Omega_B$ can have a dominant effect on quantum transport in topological materials. A modulation in $\Omega_B$ can be achieved by manipulating symmetry and band structure properties,

Sources of non-null $\Omega_B$ can also exist in real space. When an electron crosses a spatially inhomogeneous magnetic pattern, it acquires a real-space Berry phase, if the magnetic moment of the electron varies constantly along the local magnetization. The gradient of the spin orientation converts into a Berry vector potential that allows to define an emergent magnetic field. Therefore, any magnetic structure that varies smoothly in real space leads to an emergent magnetic field and in turn to a geometrical Hall effect.

Like for the spatial variation, a time variation of a spin pattern results in an emergent $E$. Among all possible magnetic spin textures, skyrmions are ideal textures to investigate these emerging electric and magnetic fields. Magnetic skyrmions with their characteristic 3D spin hedgehog typical form in materials with spiral ordering due to the Dzyaloshinskii–Moriya interaction (DMI), which can occur as result of inversion symmetry breaking, double exchange for ferromagnetic correlations or Heisenberg exchange for antiferromagnetic correlations.

In oxides like $SrRuO_3$, due to a sizable spin-orbit coupling and a non-trivial spin texture, the Berry curvature can also be strongly enhanced and modulated in sign and amplitude. This is possible due to the coexistence of a magnetic spin texture in real space and a non-trivial Berry curvature $\Omega_B$ in $k$-space in $SrRuO_3$ (Fig. 10b). Such coexistence is quite unique, but it also indicates a high complexity which requires distinct strategies for exploiting and disentangling the difference sources of Berry curvature effects. In this context, for the engineering of new devices as well as for fundamental reasons, it is challenging to evaluate how modifications of the spin texture of $SrRuO_3$ (e.g., via $V_G$-applied strain) can tune physical effects stemming from its intrinsic non-null $\Omega_B$ (see also section 2.4).

One of the current most important challenges related to $\Omega_B$ effects in SrRuO$_3$ is understanding how to differentiate and separately access real-space and $k$-space contributions to $\Omega_B$. Disentangling these two types of contributions is crucial to achieve control over their magnitudes in the Hall response and other quantum transport effects exploited for SrRuO$_3$-based quantum electronic devices. The ongoing debate on the actual existence of topological spin textures (skyrmions) in SrRuO$_3$ also fits into this wider research objective.

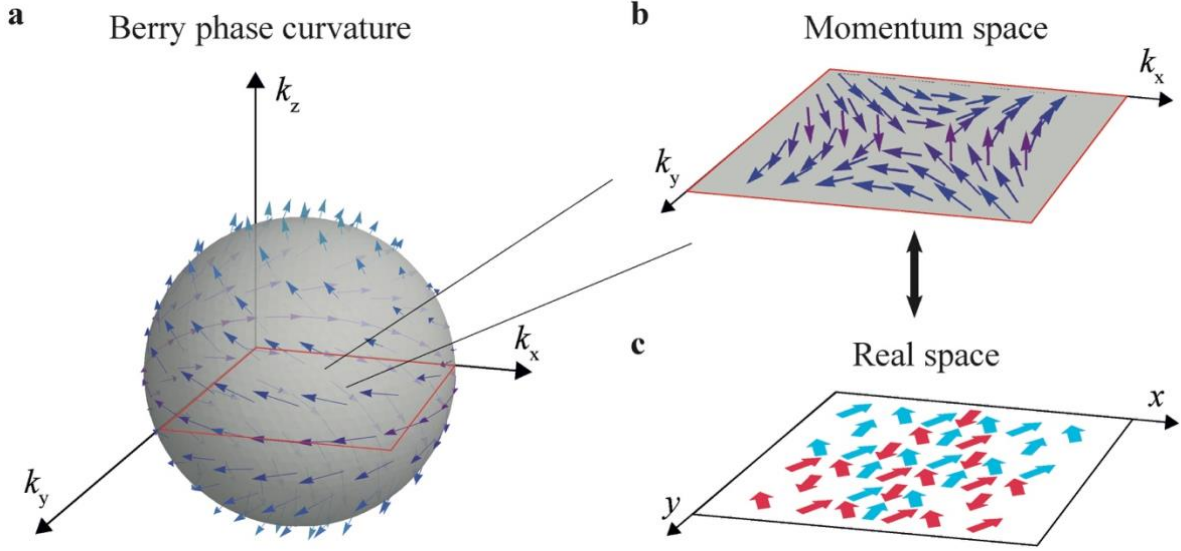

**FIG. 10. Berry effects in real and momentum space. (a) Berry curvature representation in the 3D momentum space. The vector field in the 2D momentum space shown in (b) corresponds to the azimuthal projection of the Berry curvature in (a) and it is normally correlated to a non-trivial spin texture in real space (c).**

The existence of nanometric-sized skyrmions in SrRuO$_3$ was first suggested by Matsuno and co-workers after the observation[180] of a THE manifesting through hump- and bump-like features in the $H$ field dependence of the Anomalous Hall conductance ($\sigma_{AH}$) of bilayers of SrRuO$_3$ thin films on SrIrO$_3$ (Fig. 11). The argument made by the Matsuno *et al*. to explain their observation[180] is that heavy transition-metal elements like Ir can induce DMIs in systems of reduced dimensionality like ultrathin SrRuO$_3$ films. The magnetic exchange occurring through the Ir-O bonds can induce a DMI between Ru spins in the SrRuO$_3$ ferromagnetic state which, for a suitable amplitude of DMI, would lead to the nucleation of skyrmions.

We note here that the magnetic ground-state phase diagram of SrRuO$_3$-based systems with DMIs is hard to compute theoretically because it is difficult to quantify the DMI amplitude and to use models with localized spins and short-ranged interactions in the metallic state of SrRuO$_3$. Results like those of Matsuno and co-workers[180] on the $H$-dependence of $\sigma_{AH}$ are therefore difficult to model. A similar $H$-dependence of $\sigma_{AH}$ to that first reported in ref.[180] has also been shown in other studies[181,199]. Nevertheless, features resembling a THE have also been measured for SrRuO$_3$ thin films deposited on SrTiO$_3$ without any SrIrO$_3$ or similar interface layer[200-204]. These results and the subsequent observation of bump- and hump-like features also in the $H$ variation of σ$_{AH}$ of asymmetric SrTiO$_3$/SrRuO$_3$/SrIrO$_3$ and symmetric SrIrO$_3$/SrRuO$_3$/SrIrO$_3$ trilayers[205] have led to consider alternative mechanisms to skyrmions to explain the physical origin of the hump- and bump-like features in the SrRuO$_3$ σ$_{AH}$.

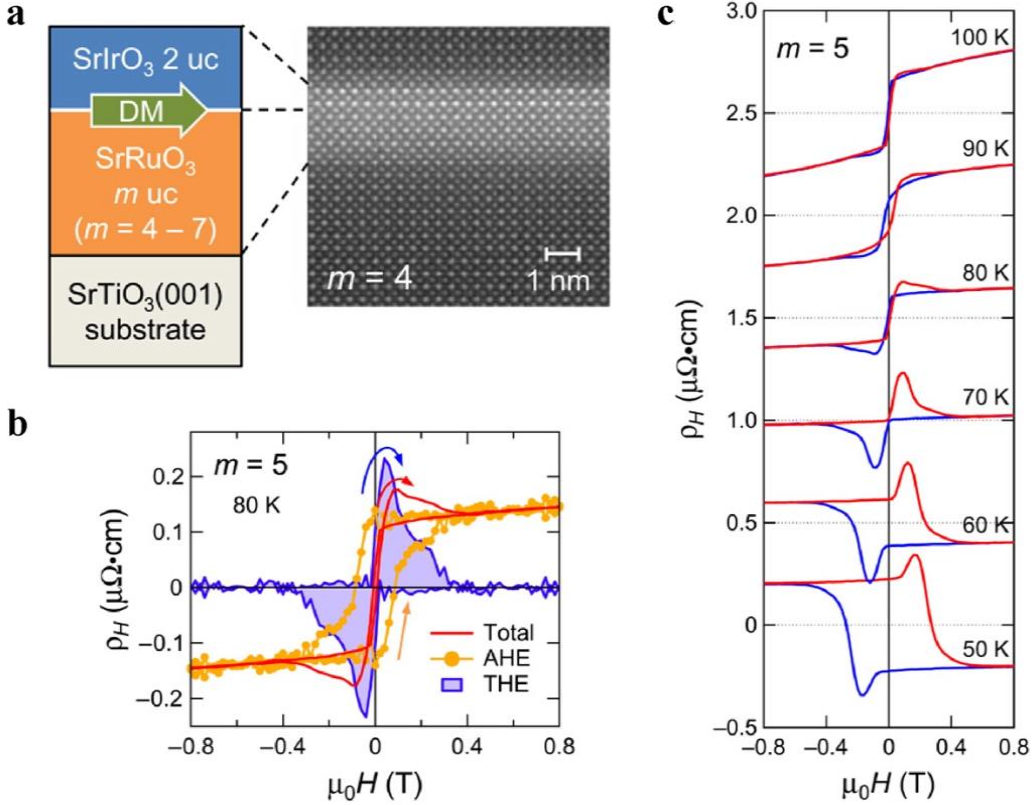

FIG. 11. Topological and anomalous Hall effect in SrRuO$_3$ systems. (a) Schematic of SrIrO$_3$/SrRuO$_3$ bilayer on SrTiO$_3$ substrate with a high-resolution transmission electron microscopy image of the SrIrO$_3$/SrRuO$_3$ interface. (b) Magnetic field dependence of the Hall resistivity ρ$_H$ of SrIrO$_3$ (2 u.c.)/SrRuO3 (5 u.c.) at $T$ = 80 K showing the contributions from the anomalous Hall and topological Hall effects with corresponding temperature evolution of ρ$_H$ for the same device shown in (c). [Figure adapted from ref. [180]].

The characteristic $T$ evolution of the σ$_{AH}$ at $H = 0$ also suggests that intrinsic contributions, in addition to real-space magnetic spin textures, must play an important role in determining the AH response of SrRuO$_3$ thin films. The sign change in σ$_{AH}$ occurring at a $T$ approximately equal to half $T_{Curie}$, and the variation in both sign and amplitude of σ$_{AH}$ when going from SrTiO$_3$/SrRuO$_3$/SrIrO$_3$ to SrIrO$_3$/SrRuO$_3$/SrIrO$_3$ trilayers[205], cannot be explained on the basis of conventional mechanisms contributing to the AHE in ferromagnetic materials like side-jump and screw-scattering contributions (Fig. 12). Also, these variations in σ$_{AH}$ cannot be accounted for only based on skyrmions, as they occur at the same $H$ values where the SrRuO$_3$ magnetization ($M$) reverses its direction in the $M(H)$ loops.

The σ$_{AH}$ variation in an applied $H$ must be also connected to the intrinsic nature of the SrRuO$_3$ electronic bands in the ultrathin limit. The low-energy electronic structure and band topology of SrRuO$_3$ is in fact characterized by topologically nontrivial spin-polarized bands at the Fermi energy (Fig. 12). These bands act as sources of non-null Berry curvature Ω$_B$ and lead to competing contributions in the AH response[205]. It is hence clear that $k$-space contributions to Ω$_B$, in addition to the real-space magnetic textures, are essential to fully understand and control the AH response of SrRuO$_3$-based systems.

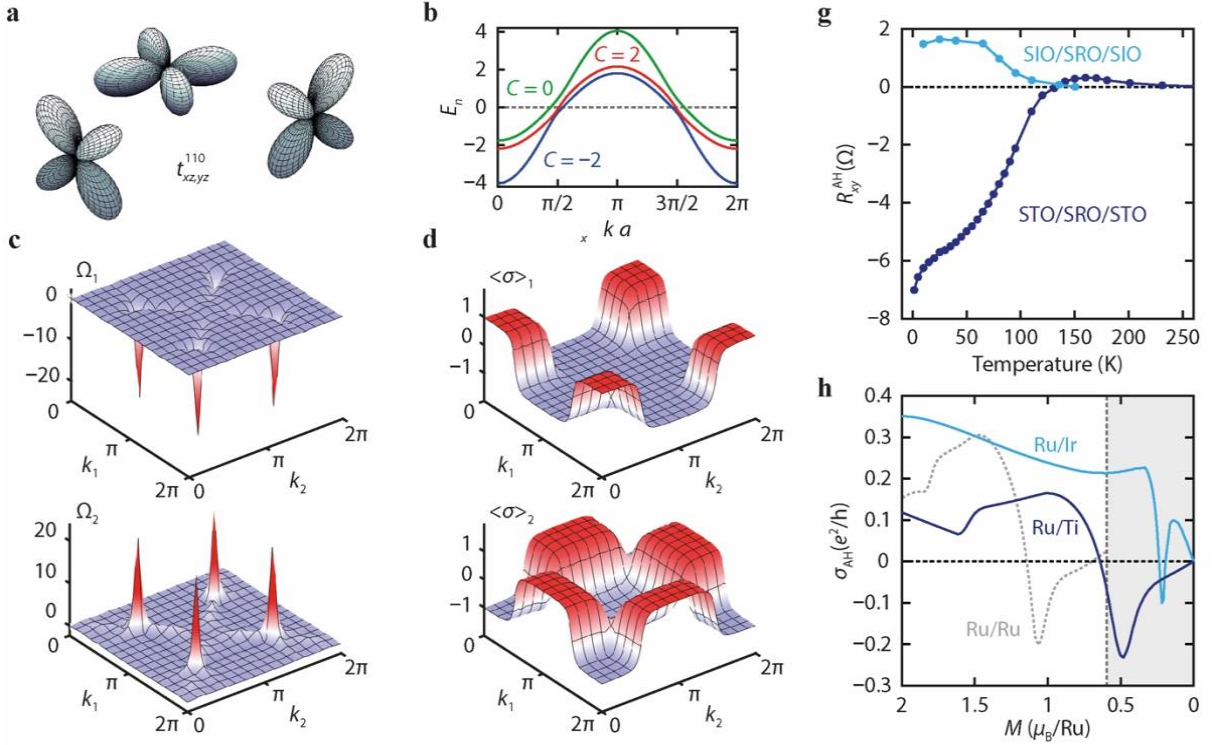

FIG. 12. Anomalous Hall effect in ultrathin SrRuO$_3$ with symmetric boundary conditions. Next-nearest-neighbor interorbital hopping (a) and dispersion of the Ru $t_{2g}$ bands along $k_x = k_y$ at a given magnetization (b). Berry curvature (c) and spin-polarization (d) associated to the topologically nontrivial Ru $t_{2g}$ bands close to the Fermi level. Temperature evolution for SrIrO$_3$/SrRuO$_3$/SrIrO$_3$ and for SrTiO$_3$/SrRuO$_3$/SrTiO$_3$ of the transverse Hall resistance $R_{xy}$ (g) and evolution of the intrinsic contribution to anomalous Hall conductivity σ$_{AH}$ for Ru/Ti, Ru/Ir and Ru/Ru as a function of the Ru magnetization (h) [Figure adapted from ref. [205]].

Apart from the above heterostructures based on SrRuO$_3$, a remarkable evolution of Ω$_B$ in $k$-space has been recently reported also for a system consisting of ultrathin SrRuO$_3$ combined with LaAlO$_3$, which is a polar wide bandgap insulator[206]. Van Thiel and co-workers have shown[206] that the synthesis of RuO$_2$-terminated SrRuO$_3$ ultrathin films interfaced with LaAlO$_3$ results in levels of charge doping of SrRuO$_3$ that go well beyond those obtainable with electrostatic gating. The high doping results in a pronounced profile with excess electron density along the growth axis of the SrRuO$_3$ thin film. In the ultrathin limit of SrRuO$_3$, the doping-induced electronic charge reconstruction leads in turn to a variation of the $\Omega_B$ sign in $k$-space, which manifests experimentally as a variation in the $\sigma_{AH}$ sign[206].

The theoretical analysis carried out in ref.[206] identifies the charge pinning at the SrRuO$_3$/LaAlO$_3$ interface and the resulting inversion symmetry breaking as the dominant mechanisms responsible for the reconstruction of $\Omega_B$ in $k$-space. This implies that the change in $\Omega_B$ sign is a consequence of a topological-like transition in $k$-space other than of a change in the electronic band occupation. The results of this work[206] suggest that electronic charge reconstruction can be used in the future as an effective tool to manipulate $\Omega_B$ and correlated topological transitions in SrRuO$_3$, which in turn affect measurable quantities like $\sigma_{AH}$.

Based on the above consideration, it is evident that SrRuO$_3$ represents a material with potential coexistence of $k$- and real-space Berry effects, whose origins and characteristic scales are completely distinct. A remarkable aspect of this coexistence is that topological configurations in real and $k$-space occur only for specific regions of the phase diagram as a function of parameters such as $T$, $H$ and electron filling.

Apart from mapping the parameters' space to determine the configurations with a dominant real- or $k$-space character of $\Omega_B$ in SrRuO$_3$, another future challenge is to differentiate configurations based on real-space topological spin textures from those with a non-trivial topology in $k$-space. To address all these questions, we suggest two possible experiments.

Our first proposal is sketched in Fig. 13 and exploits the spin dependence of $\sigma_{AH}$ in SrRuO$_3$ in its ferromagnetic state. The key point here is to evaluate the spin content of the AH voltage measured across a SrRuO$_3$ Hall bar. To do this, a spin-polarized current can be injected into SrRuO$_3$ (e.g., through a half-metal ferromagnet coupled to SrRuO$_3$) and the resulting AH voltage should be detected with ferromagnetic electrodes. This should be done for different configurations, where the magnetization is switched from parallel to antiparallel with respect to the SrRuO$_3$ magnetization or from oriented along the SrRuO$_3$ easy axis or along the SrRuO$_3$ hard magnetic axis. The as-measured transverse Hall signal would contain information about transport processes conserving spin and can be compared (in sign and amplitude) to another transverse Hall signal measured on the same Hall bar with normal-metal electrodes (Fig. 13). The comparison would allow to understand whether the transverse voltage is due to a $\Omega_B$ dominated by spin-conserving processes (related to $k$-space topological contributions) or by non-conserving spin scattering processes (related to real-space topological contributions).

Our second proposal to understand the dominant contributions to $\Omega_B$ in SrRuO$_3$ is based on the design of heterostructures where SrRuO$_3$ is interfaced with a superconducting material. As discussed in detail in section 2.4, we expect that the interplay of magnetic states having a non-trivial $\Omega_B$ in real or *k*-space with a superconductor would allow to distinguish between the two types of topological contributions.

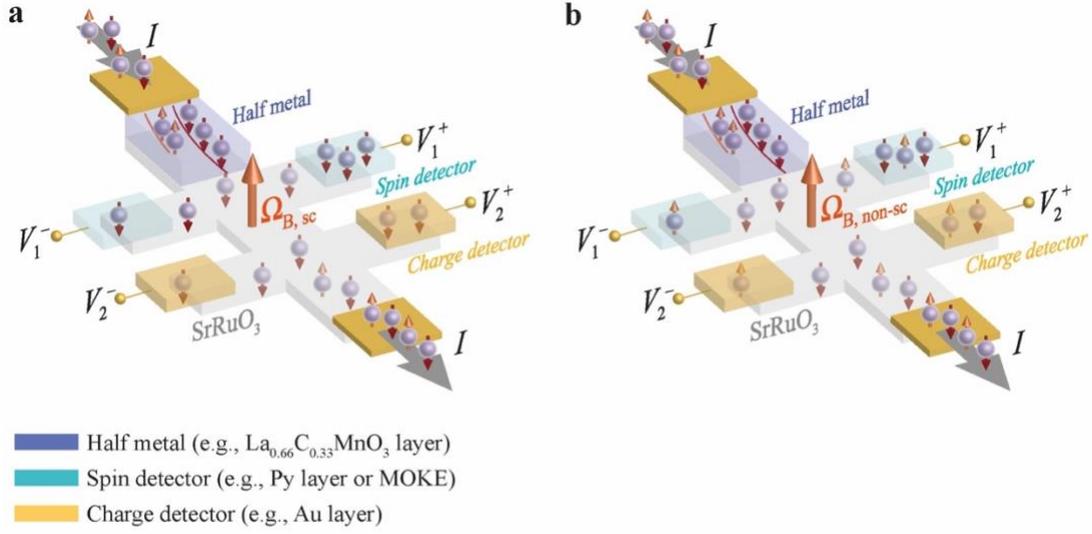

**FIG. 13. Berrytronic device to determine the nature of Berry effects in SrRuO$_3$.** The device consists of a SrRuO$_3$ Hall bar, where a fully spin-polarized current, which is generated by current injection through a half-metal, induces a transverse voltage detected with both spin-sensitive (e.g., ferromagnet) and charge-sensitive (e.g., normal metal) contacts. If *k*-space contributions are dominant, the Berry curvature is expected to be mostly spin-conserving and the transverse voltage measured with the spin detectors has a sign dependent on the half metal magnetization and that can be opposite to the voltage signal measured via charge detectors (a). The scenario in (a) is opposite to that shown in (b), where dominant real-space contributions make the Berry curvature mostly not spin-conserving and the two transverse voltage signals have always the same sign.

## 2.4 Topological superconductivity and superconducting berrytronics

Due to its good lattice matching with other oxide perovskites including high-temperature superconductors (HTSs) like YBCO, SrRuO$_3$ has been already studied in a variety of superconducting devices such as Josephson junctions[26-28,207,208] (JJs) and superconducting spin valves[209]. As a result of its good lattice matching with YBCO and thermal stability, it has also been shown that SrRuO$_3$ can also be used as buffer layer to improve the performance of HTS coatings[210] and to boost their superconducting critical current ($I_c$) density.

Several groups have characterised the superconducting properties of JJs with SrRuO$_3$ as weak link including YBCO/SrRuO$_3$/YBCO JJs (refs.[26, 28,207]) and hybrid metal/metal-oxide JJs like Nb/Au/La$_{0.7}$Sr$_{0.3}$MnO$_3$/SrRuO$_3$/YBCO (ref. [208]). Most of these experiments and independent low-*T* STS measurements on SrRuO$_3$/YBCO bilayers[142] suggest that the superconducting order parameter can penetrate into SrRuO$_3$ over a depth larger than 20 nm at 4.2 K (refs. [26,27,142,207]), which is an order of magnitude larger than the typical superconducting coherence length $\xi_F$ in a strong ferromagnetic metals like Ni or Co (~ 1-2 nm; refs. [211,212]). This long-ranged proximity effect has been ascribed to crossed Andreev reflections taking place near domain walls at the $SrRuO_3$/YBCO interface[142] or to resonant tunnelling of quasiparticles through an oxygen-depleted layer forming at the $SrRuO_3$/YBCO interface[207]. It should be noted, however, than in hybrid metal/metal-oxide Nb/Au/$La_{0.7}Sr_{0.3}MnO_3$/$SrRuO_3$/YBCO JJs a long-ranged proximity effect is only observed when both ferromagnets ($La_{0.7}Sr_{0.3}MnO_3$ and $SrRuO_3$) are present[208]. The authors of ref.[208] relate this long-ranged proximity to the formation of spin-polarised (spin-triplet) Cooper pairs generated by a non-collinear magnetization[213] in the LSMO/SRO system.

The generation of spin-triplet pairs in superconductor/ferromagnet (S/F) hybrids has been investigated by several groups due to its potential for the realization of spintronic devices operating in the superconducting state with very low energy dissipation. Within this research field, known as superconducting spintronics[214], $SrRuO_3$ has been shown to be promising for the realization of $Sr_2RuO_4$/$SrRuO_3$ (S/F) devices.

Anwar et al. have demonstrated[215] that the growth by PLD of $SrRuO_3$ at high $T$ (~ 600 °C) on single-crystal $Sr_2RuO_4$ substrates results in a reorientation of the $RuO_6$ octahedra on the $Sr_2RuO_4$ surface leading to proximity-induced superconductivity in $SrRuO_3$. This is an important result for applications because bare $Sr_2RuO_4$ single crystals have an insulating surface due to atomic reconstructions[216] – which makes it difficult to study the superconducting proximity effect between the superconductor $Sr_2RuO_4$ and other materials. According to Anwar and co-workers, the PLD growth of $SrRuO_3$ on $Sr_2RuO_4$, restores metallic behaviour at the $SrRuO_3$/$Sr_2RuO_4$ interface and it allows to measure proximity-induced superconductivity in $SrRuO_3$ over a $\xi_F$ of ~ 9 nm (ref. [217]). In addition to the long $\xi_F$, which is comparable to that reported in YBCO/$SrRuO_3$ systems[26,27,142,207], Anwar et al. also studied the proximity-induced superconducting gap in $SrRuO_3$ by fabricating Au/$SrTiO_3$/$SrRuO_3$/$Sr_2RuO_4$ tunnel junctions[218]. The shape and $T$-evolution of gap features in the differential conductance $dI/dV$ of these junctions show an unusual behavior which the authors reconcile with an anisotropic superconducting gap induced in $SrRuO_3$ with $p$-wave or $d$-wave symmetry[218]. It is worth noting that the interplay between different mechanisms including orbital loop current magnetism recently discovered[219] at the $Sr_2RuO_4$ surface and inverse proximity[220] makes the $Sr_2RuO_4$/$SrRuO_3$ interface a complex system to study and that can indeed host spin-triplet and other unconventional superconducting states.

The study of the interplay of Berry effects in SrRuO$_3$ with conventional or unconventional superconductors represents an unexplored line of research, which can lead to the discovery of topologically protected superconducting states for quantum electronics.

We first discuss here the topological phases that may arise if SrRuO$_3$ is coupled to another spin-singlet superconductor. The first case that we consider refers to the superconducting proximity between a conventional spin-singlet superconductor and SrRuO$_3$ acting as a topological metal with uniform magnetization. This assumption is supported by the fact that SrRuO$_3$ has electronic bands with a non-trivial topological character and non-null Chern number. The superconducting proximity between a topological metal like SrRuO$_3$, in a state with uniform magnetization below $T_{Curie}$, and a spin-singlet *s*-wave superconductor can result in the emergence of topological superconductivity with topologically protected nodes at the Fermi level. This topological superconducting state induced in SrRuO$_3$ by proximity effect is analogous to that deriving from the formation of pairing correlations between electronic states of bands with nontrivial topological structure and marked by a gauge flux distribution in *k*-space[221]. Unlike for the typical superconducting proximity effect, where the structure of the superconducting gap in *k*-space is related to the attractive potential responsible for the formation of pairing correlations, electronic pairing in topological bands is determined by the topological character of the normal-state electronic wavefunctions. In proximitized 2D vdW heterostructures, for example, the vorticity of the superconducting gap in *k*-space is set by the vorticity associated to the Chern number for each of the spin of the two electrons forming a Cooper pair[221]. This also suggests that a non-trivial monopole density can arise in a topological superconducting state.

The topological superconductivity which we envision in these spin-singlet s-wave superconductor/SrRuO$_3$ systems should be robust against disorder, and it can lead to anomalies in both charge (Josephson effect) and heat transport in superconducting devices. Also, the way in which the vorticity of electrons pairing in *k*-space can affect the formation of Andreev bounds states in the proximitized region close to the interface between a spin-singlet *s*-wave superconductor and a ferromagnet like SrRuO$_3$ represents an open question still yet to address. We expect that a very good electronic matching is needed at the interface between SrRuO$_3$ and a conventional *s*-wave superconductor to trigger topological superconductivity. For this reason, the epitaxial growth of a metal-oxide superconductor (e.g., LiTi$_2$O$_4$) with a spin-singlet *s*-wave order parameter onto SrRuO$_3$ would be ideal to meet this requirement.

Another interesting system which can be explored in the future for the realization of a topological superconducting phase consists of SrRuO$_3$ proximity coupled to a nodal *d*-wave superconductor like YBCO, as shown in Figs. 14a and b. In such case, the combination of induced spin-triplet pairs at the YBCO/SrRuO$_3$ interface with inversion symmetry breaking can gap the otherwise nodal (i.e., gapless) excitations in YBCO and turn YBCO into a topological superconductor[222]. This type of topological phase engineering in the superconducting state has been proposed[223] for systems consisting of a HTS in the presence of an applied magnetic field or of heavy fermion superlattices, where mixed spin-triplet pairing is generated through inversion symmetry breaking. A challenging aspect to achieve topological superconductivity in SrRuO$_3$/YBCO is to control the strength of the inversion symmetry breaking, the magnetic anisotropy and the degree of inverse proximity at the SrRuO$_3$/YBCO interface. Here, we point out that the lack of observation of the topological phases to date in the SrRuO$_3$/YBCO may be due to the absence of a significant contribution of the spin-triplet order parameter in a spin-singlet superconductor. To overcome this issue and realize topological superconductivity, an important ingredient, which has not been fully exploited to date, could be given by an increase in the atomic spin-orbit coupling at the SrRuO$_3$/YBCO interface. Indeed, one can employ an interlayer between YBCO and SrRuO$_3$ that is marked by a strong spin-orbit coupling to enhance the spin-triplet pairing component in the *d*-wave superconductor. For instance, an heterostructure of the type SrRuO$_3$/SrIrO$_3$/YBCO with few layers of SrIrO$_3$ at the interface of YBCO can be a viable route to increase the density of spin-triplet pairs and in turn obtain a topological superconductor. Also, since the spin-triplet order parameter has a vectorial character described by the so called d-vector, another challenging aspect is the control of the d-vector orientation due to its coupling with the magnetization[224] at the interface and in the presence of inversion symmetry breaking. Furthermore, inverse proximity effects with the leakage of the spin polarization from SrRuO$_3$ into YBCO can play a relevant role in reconstructing the d-vector[220] and in turn to achieve a topological superconducting state – which can be switched on/off depending on the magnetic anisotropy and degree of electronic matching at the interface. In the topological superconducting phase, we expect the appearance of chiral currents at the edge of the superconductor, which can be visualized by means of magnetic imaging technique like scanning SQUID magnetometry. Moreover, the charge and spin conductance will be affected by the presence of topological modes in a way that will be different from the case of tunneling into a pure nodal *d*-wave superconductor.

An additional path that we foresee for the realization of topological superconductivity in superconducting heterostructures based on SrRuO$_3$ stems from the non-collinear magnetic spin textures (e.g., skyrmions) which have been suggested to nucleate in ultrathin SrRuO$_3$ at its coercive field or in heterostructures[180] combining SrRuO$_3$ with a high-spin orbit coupling material like SrIrO$_3$ (see also section 2.3). The proximity effect between a conventional spin-singlet superconductor and a non-collinear magnetic spin texture (Fig. 14c) can be exploited to convert spin-singlet pairs into chiral or helical spin-triplet pairs. This physical scenario is inspired by the theoretical finding that an s-wave superconductor can be turned into a p-wave superconductor, if it is interfaced with a semiconductor with large Rashba spin-orbit interaction, under the assumption that a source of time-reversal symmetry breaking (e.g., a magnetic exchange field) is also present[225,226]. Fabricating this type of devices sketched in Fig. 14b, however, requires achieving a systematic control over the generation of skyrmions in SrRuO$_3$-based systems and then performing systematic studies on their coupling to superconductors.

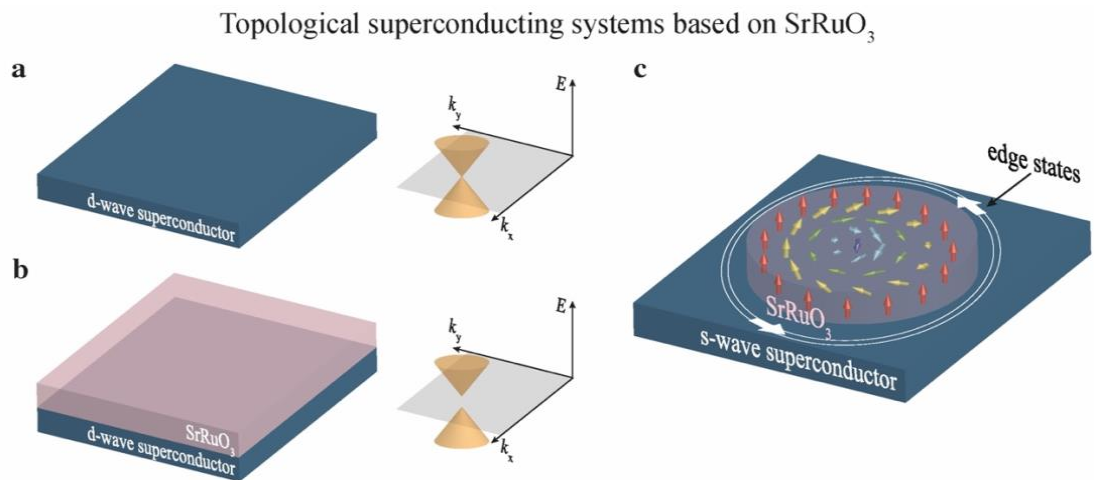

**FIG. 14. SrRuO$_3$-based system for realization of topological superconductivity. Illustration of a superconducting heterostructure consisting of a *d*-wave superconductor (e.g., YBCO) with nodal gapless density of states (a) and realization of a topological superconducting state in YBCO with gapped density of states due to a combination of inverse proximity with SrRuO$_3$, spin-orbit coupling and inversion symmetry breaking (b). Schematic of another system for the realization of topological superconductivity (c) consisting of an s-wave superconductor in proximity coupling with a non-collinear magnetic spin texture (e.g., a skyrmion) in SrRuO$_3$. The topological states forming at the boundary of the non-collinear magnetic region are chiral and give rise to a spontaneous current flowing along the edge.**

Since a rotating magnetic field is equivalent, from the point of view of conversion of spin-singlets into spin-triplets, to the combination of Rashba spin-orbit coupling with an applied homogenous magnetic field, one can engineer quasi-1D topological superconductors with magnetic spin textures, or alternatively with antiferromagnetism or ferromagnetism in the presence of external currents and Zeeman fields. A magnetic helix crystal hence represents a suitable system to realize topological superconductivity when coupled to a conventional superconductor, since a magnetic helix can simultaneously generate spin-orbit coupling (due to inversion symmetry breaking) and a magnetic exchange field. While a magnetic helix is sufficient to induce a topological superconducting state, to achieve a strong topologically protected state in a number of dimensions greater than one, however, it is necessary that the magnetic spin texture winds in all direction. As a result, whilst a magnetic helix coupled to a conventional superconductor can induce spinless *p*-wave pairing in 1D, a spin skyrmion is necessary to get an effective spinless chiral *p*+i*p* topological superconductor in 2D. Evidence for topological superconductivity stabilized by non-trivial magnetic spin textures has been recently demonstrated in various materials platforms consisting of magnetic atoms/clusters deposited on a superconductor surface or of superlattices hosting chiral magnetic textures[227-230]. One of the challenges to address in the future to achieve topological superconductivity from the proximity effect between a superconductor and magnetic skyrmions in $SrRuO_3$ is to control the mutual competition between the magnetic and superconducting order parameters and to determine the best magnetic spin texture for the realization of topological superconductivity[231].

A magnetic skyrmion in $SrRuO_3$ can also trigger formation of vortices into a superconductor coupled to $SrRuO_3$. The spin polarization of the skyrmion combined with the spin-orbit coupling can induce a charge current at the superconductor/$SrRuO_3$ interface. An important challenge here is to differentiate between effects genuinely induced by the exchange coupling between the skyrmions in $SrRuO_3$ and the superconductor from those instead merely related to the magnetic stray fields. It should be noted that exotic spin-polarized quasiparticle states can also form in these topological superconducting phases – these quasiparticle states can be exploited for low-dissipation spin transport in the superconducting state[232].

Although the complexity of the superconducting topological phases based on $SrRuO_3$/superconductor hybrids is very high, there are several degrees of freedom that can be exploited to control these phases including the type of magnetic spin texture in $SrRuO_3$ triggering them, their shape, and the strength of their coupling between the spin texture and the superconducting condensate. Deviations of the magnetic spin texture from a magnetic helix, for example, can induce different types of topological superconductivity due to changes in the corresponding spatial distribution of the magnetic moments. For an inhomogeneous magnetic helix, for example, topological domains may form inside the magnetic material along with topologically protected modes nucleating at the domain walls[233]. This suggests that control over topological superconducting phases can be achieved, for example, by engineering domains with inequivalent non-collinear magnetic spin texture. Local spectroscopy techniques can be used to resolve the spatial profile of the magnetic texture. We expect that variations in the magnetic spin textures are likely to occur in $SrRuO_3$ and $SrRuO_3$-based heterostructures due to the itinerant ferromagnetism of $SrRuO_3$ and to nonuniform stray fields.

In addition to the generation of topological superconductivity, we foresee another important application of $SrRuO_3$, which relies on using its Berry curvature as mechanism for spin-triplet generation in superconducting spintronic devices. The possibility that magnetic materials with non-null Berry curvature can be used to convert spin-singlet pairs into spin-triplet pairs has been proposed in a recent study[234], where the authors have reported long-ranged Josephson coupling (up to ~ 160 nm) between two Nb electrodes separated by the chiral antiferromagnet $Mn_3Ge$. When the antiferromagnet $Mn_3Ge$, which has non-null Berry curvature, is replaced by another antiferromagnet (IrMn) with trivial spin texture and null Berry curvature, no long-ranged currents due to spin-triplet pairs is observed[234].

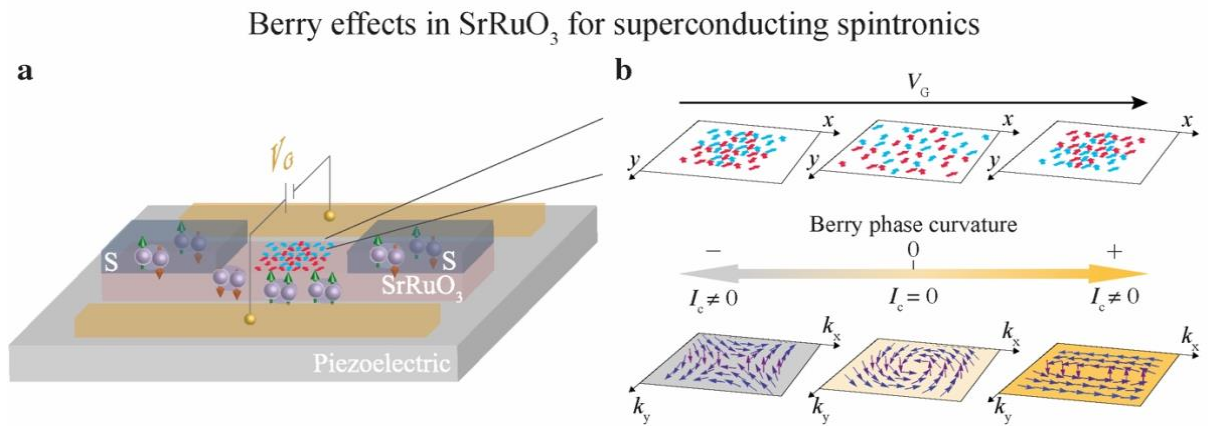

**FIG. 15. Superconducting spintronics with $SrRuO_3$ exploiting Berry effects. Illustration of a device for reversible control over spin-triplet generation induced by the non-null Berry curvature of a $SrRuO_3$ weak link separating two superconducting (S) electrodes (a). The application of a gate voltage $V_G$ to a piezoelectric coupled to $SrRuO_3$ induces strain-driven modifications in its real-space spin texture, which in turn result in variations (in sign and amplitude) of the $SrRuO_3$ Berry curvature (b). The modulation of the Berry curvature leads to changes in the amplitude of the spin-triplet critical current $I_c$ flowing between the two S electrodes, which switches between null (small) and non-null (large) values thus realizing the equivalent of a superconducting switch.**

Compared to the conventional mechanism used to date in superconducting spintronics for spin-triplet generation, which consists in coupling of a spin-singlet superconductor to a ferromagnet with an inhomogeneous magnetization[235-236] (or a to stack of ferromagnets with non-collinear magnetizations[237]), using the Berry curvature as alternative mechanism for spin-triplet generation offers several advantages for applications.

In materials like SrRuO$_3$ due to its sizable spin-orbit coupling and a non-trivial spin-texture, the Berry curvature can be strongly enhanced and modulated (in sign and amplitude) due to the correlation existing between magnetic spin texture in real space and Berry curvature in $k$-space. This also implies that, in Josephson junctions where SrRuO$_3$ is used as weak link between two superconducting electrodes, changing the magnetic spin texture of SrRuO$_3$ in real space (e.g., via $V_G$-driven strain through a piezoelectric coupled to SrRuO$_3$) can in turn affect its Berry curvature in $k$-space and therefore reversibly enhance or suppress the spin-triplet channel in SrRuO$_3$ (Fig. 15). If the two superconducting electrodes are separated by a distance larger than the spin-singlet coherence length, switching on/off the long-ranged spin-triplet channel in SrRuO$_3$, can turn the SrRuO$_3$ weak link from resistive (triplets off) to superconducting (triplet on). This type of superconducting device would act as a switch and represent the first superconducting spintronic device with full electrical control of its state.

Voltage-driven devices would offer many advantages compared to existing superconducting spintronic devices, whose state is currently controlled by switching the ferromagnet's magnetization from homogeneous (triplets off) to inhomogeneous (triplets on) via an applied magnetic field. Superconducting devices with magnetic control of their logic state are in fact more sensitive to environmental noise, less scalable and less energy efficient than equivalent devices whose logic states is controlled electrically.

We also note that the Berry curvature per se acts for electrons as the equivalent of a magnetic field. Therefore, in addition to variations in the Berry curvature of SrRuO$_3$ induced by voltage-driven strain, one may fabricate superconducting spintronic devices where the combination of spin-polarization (in SrRuO$_3$ itself or in another oxide ferromagnet coupled to SrRuO$_3$) and Berry curvature in SrRuO$_3$ can be used for the generation of spin-triplet pairs for superconducting spintronics.

## 2.5 Summary and outlook

In this Research Update, we have given an overview of some of the most recent work done on SrRuO$_3$ which holds promising potential for the development of novel electronic (conventional and quantum) applications. We have first discussed the main physical properties of SrRuO$_3$, which have kept the interest in material always very high over the past 60 years, and the most recent advances in recent techniques for the fabrication of high-quality SrRuO$_3$ with high reproducibility and over large scales. We have then explained the structural parameters and experimental tools which previous studies have demonstrated to be useful to control specific SrRuO$_3$ properties. To illustrate how properties change with dimensionality and confinement, which is relevant for quantum applications based on SrRuO$_3$, we have also reviewed progress recently made on SrRuO$_3$ structures with dimensionality lower than 3D.

In the second part of this manuscript, we have discussed how, thanks to its rich physics, SrRuO$_3$ represents a material platform with great potential for the realization of electronic devices not only useful for conventional electronics, but also for emerging quantum electronics. In this section of our Research Update, we have not only limited ourselves to review recent progress made on SrRuO$_3$ devices, but also taken some personal perspectives on future research directions which can bring new insights into effects recently discovered in SrRuO$_3$. We have also proposed devices never realized to date both for conventional and quantum electronics and sketched possible layouts useful for their realization. From this point of view, we hope that this manuscript will inspire the research community to perform new investigations on some of the SrRuO$_3$ heterostructures and devices that we propose.

For the specific application of SrRuO$_3$ for conventional electronics, we have discussed two of most promising applications where SrRuO$_3$ devices can offer a competitive advantage over existing ones. These two applications concern the realization of racetrack memories based on domain wall motion and spin-orbit-torque memories. In addition to large-scale production and reproducibility in their properties, which are essential requirements to meet for applications, other materials challenges must be faced for the realization of such SrRuO$_3$ devices. These challenges include obtaining reversible control over the strength of the spin-orbit coupling in SrRuO$_3$ (e.g., via modulation in the tilting of the RuO$_6$ octahedra), quantifying the width of SrRuO$_3$ domain walls and achieving their manipulation under current injection.

Within the field of conventional electronics, we have also outlined that the very recent realization of ultrathin freestanding SrRuO$_3$ membranes can pave the way for the fabrication of NEMS devices and sensors with unprecedently high figures out merit. The fabrication of SrRuO$_3$ membranes with optimal properties and the testing of their reliability over several operation cycles remain key materials challenges for the future development of these devices.

In the field of the quantum electronics, future applications will certainly stem from the interplay between different mechanisms and quantum effects in SrRuO$_3$. It is currently well-established that SrRuO$_3$ becomes a very rich quantum system close to the 2D limit and when interfaced to other materials. We have explained that the possibility to couple different quantum orderings and phases at SrRuO$_3$ interfaces and to tailor the confinement potential in the ultrathin limit provides novel paths for the generation, control, and manipulation of electronic states with nontrivial Berry curvature and topological properties.

As we have discussed in the manuscript section on quantum applications, the interplay of Berry curvature and non-trivial topological states with superconductivity paves the way for the testing and fabrication of a new quantum electronic devices. The devices which we propose exploit quantum effects provided by the Berry phase of $SrRuO_3$ in both real and momentum space. Being able to differentiate between momentum-space (spin-conserving) and real-space (non-spin-conserving) contributions to the $SrRuO_3$ Berry curvature remains one of the most important challenges to realize berrytronic devices on $SrRuO_3$. Also, the realization of superconducting systems where the $SrRuO_3$ Berry curvature acts as a mechanism for spin-triplet generation and it can be reversibly manipulated (in sign and amplitude), can lead to the realization of the first class of superconducting spintronic devices with full electrical control of their state. A hallmark feature of the quantum devices that we envision is their tuneability achieved through control of magneto-orbital effects, strain, and interfacing of $SrRuO_3$. This area is not yet fully explored and calls for significant research efforts, particularly in materials science, to master quantum transport properties and coherent effects arising from the $SrRuO_3$ electronic and magnetic states.

In addition to the promising applications described above in the manuscript, there are other research directions with great potential for the discovery of novel effects in $SrRuO_3$ and the consequent development of devices relying on the same effects.

One of these new research directions concerns the study of quantum effects related to the geometric properties of the electronic structure of $SrRuO_3$. We have already outlined that $SrRuO_3$ is characterized by a Berry curvature that has sources both in real and momentum space and that can be tuned through various parameters including dimensionality, strength of the magnetization, inversion symmetry breaking, interfacing with other materials. We expect that exciting discoveries can be made in future studies on Berry effects in ultrathin $SrRuO_3$ films. This is because, for ultrathin $SrRuO_3$ films close to the one-unit-cell-thick limit, sources of Berry curvature in real space can be nucleated at the $SrRuO_3$ film surface or at the interface with another material inducing inversion symmetry breaking. These systems can trigger the formation of distinct magnetic patterns, which may act as source of nonvanishing Berry curvature whilst retaining a topological character. Also, ultrathin $SrRuO_3$ films can be coupled to oxides with properties that can also affect the Berry curvature like strong spin-orbit coupling, large structural mismatch, polar interface, and superconductivity. Experimental evidence for magnetic patterns at the surface or interface of ultrathin $SrRuO_3$ films is still missing. The hurdles in the identification of these magnetic patterns also suggest that SrRuO$_3$ is a unique platform to develop and test new experimental probes and setups suitable to detect such non-trivial magnetic patterns. It is worth noting that the connection between magnetic patterns and Berry curvatures is per se very complex and it will require dedicated studies to gain further insights into it. Even a simple uniform magnetic domain has topological electronic bands in momentum space, with electronic charges that can be controlled via an applied *E* or strain and that depend on the strength of magnetism and Rashba spin-orbit coupling.

SrRuO$_3$ also represents an ideal platform to investigate emergent phenomena in correlated topological metals. From this point of view, we believe that future studies on topological magnetic effects in high-electron density conditions can be carried out using SrRuO$_3$ other than semimetallic materials or materials with low-carrier density. This proposed line of research can lead to the discovery of new magnetotransport effects deriving from the combination of the high sensitivity of strongly correlated electron systems (as they undergo phase reconstructions) with phase transitions induced by small changes in an external perturbation. In addition, the interplay between Coulomb interaction, spin-orbit coupling and crystal field potentials in SrRuO$_3$ can also trigger magnetotransport effects that are scalable in space and controllable in the time domain. This is another exciting research direction that remains to date unexplored.

The orbital quantum degrees of freedom are another important feature of SrRuO$_3$, whose potential has not been fully explored to date. It is well-established that SrRuO$_3$ is a multi-orbital ferromagnet and that the orbital character of its electronic states at the Fermi level can be modified via an applied *E*, strain, or geometric design. Studies aiming at controlling orbital effects in SrRuO$_3$ under external stimuli, however, remain still at their infancy. This suggests that SrRuO$_3$ offers an enormous potential for the discovery of orbital effects and the development of orbitronic devices. We believe that future studies targeting specifically the control over the orbital quantum degree of freedom in SrRuO$_3$ may lead to the detection of large orbital Hall effects or orbital selective anomalous Hall effects. The discovery of orbital Hall effects can set the basis for low-consumption quantum spin orbitronic[238]. This perspective is particularly relevant in SrRuO$_3$ structures with reduced dimensionality, where confinement and inversion symmetry breaking can be used to control the orbital population and the orbital angular momentum at the Fermi level.

Another major research route that can lead to important fundamental discoveries is the study of SrRuO$_3$-based heterostructures combining the magnetic properties of SrRuO$_3$ with superconductivity. In section 2.4 we have proposed several SrRuO$_3$-based superconducting devices which can be tested and that can lead to a paradigm shift in the field of superconducting spintronics. Once again, the realization of topological superconducting phases with Cooper pairs having non vanishing spin and orbital angular momenta (i.e., spin- and orbital- triplet pairs) may be easier for ultrathin SrRuO$_3$ films with topological electronic bands. As discussed in section 2.4, one of the major material challenges to achieve topological superconductivity, however, is to obtain a high interface quality between SrRuO$_3$ and another superconductor.

The successful integration of Berry curvature effects with superconducting spintronic elements can also facilitate the developments of electronic devices where spin Hall effects or anomalous Hall effects can be employed to control the superconducting supercurrent and vice versa. If these novel superconducting berrytronic devices were realized, they would represent a huge boost for low-consumption quantum electronics.

More research studies should also be carried out to clarify the physical mechanisms behind phenomena recently discovered in SrRuO$_3$ like the Hall crystal effect[239], phonon-driven magnetic exchange[240], and magnetic domain manipulation[241].

An obvious drawback for device applications of SrRuO$_3$ in the field of conventional spintronics is the fact that the $T_{Curie}$ of SrRuO$_3$ is below room $T$. A critical challenge is therefore to find ways to increase the $T_{Curie}$ of SrRuO$_3$. A route that could be tested for this would consist in developing a suitable geometric design to modify the bandwidth of the electronic bands and in turn enhance the density of states of SrRuO$_3$ at the Fermi level. An alternative to such approach would consist in employing substitutional transition metal elements to increase the magnetic moment strength in SrRuO$_3$. This could be carried out, for example, using Fe or Mn as substitutional dopants for Ru in SrRuO$_3$.

Future work on SrRuO$_3$ heterostructures can also lead to great technological advancements, especially after freestanding SrRuO$_3$ nanomembranes are fully integrated into them[174]. The study of the effects of geometric parameters related to the large curvature of nanomembranes on SrRuO$_3$ properties is still at its infancy. It is very likely, however, that studies on the topic may lead to the discovery of magnetic and topological Hall effects that are fully geometrically driven and that can have an impact on novel quantum electronic devices.

**Acknowledgments**
The authors wish to thank Prof. A. Caviglia and Dr. A. Vecchione for scientific discussion and comments about the manuscript.
A.D.B. acknowledges funding from the Humboldt Foundation in the framework of a Sofja Kovalevskaja grant and funding from the Zukunftskolleg at the University of Konstanz. M. C. acknowledges support from the MIUR-PRIN grant 20177SL7HC for the project "Two-dimensional Oxides Platforms for SPINorbitronics nanotechnology (TOPSPIN)." A. D. B. and M.C. also acknowledge support by the EU's Horizon 2020 research and innovation program under Grant Agreement No. 964398 (SUPERGATE).

**Data availability**
Data available on request from the authors.